\documentclass[pre,aps,twocolumn,showpacs]{revtex4}
\usepackage{amsmath,amssymb,graphicx}
\usepackage{epsfig}
\usepackage{textcomp}
\usepackage{color}

\begin{document}
\title{Extinction rates of established spatial populations}
\author{Baruch Meerson}
\affiliation{Racah Institute of Physics, Hebrew University of
Jerusalem, Jerusalem 91904, Israel}
\author{Pavel V. Sasorov}
\affiliation{Institute of Theoretical and
Experimental Physics, Moscow 117218, Russia}
\begin{abstract}
This paper deals with extinction of an isolated population caused by  intrinsic noise. We model the population dynamics  in a ``refuge" as a Markov process which involves births and deaths on discrete lattice sites and random migrations between neighboring sites. In extinction scenario I
the zero population size is a repelling fixed point of the on-site deterministic dynamics. In extinction scenario II the zero population size is an attracting fixed point, corresponding to what is known in ecology as Allee effect. Assuming a large population size, we develop WKB (Wentzel-Kramers-Brillouin) approximation to the master equation. The resulting  Hamilton's equations encode the most probable path of the population toward extinction and the mean time to extinction. In the fast-migration limit these equations coincide, up to a canonical transformation, with those obtained, in a different way, by Elgart and Kamenev (2004). We classify possible regimes of population extinction with and without an Allee effect and for different types of refuge and solve several examples analytically and numerically. For a very strong Allee effect the extinction problem can be mapped into the over-damped limit of theory of homogeneous nucleation due to Langer (1969). In this regime, and for very long systems, we predict an optimal refuge size that maximizes the mean time to extinction.

\end{abstract}
\maketitle
\section{Introduction}
\label{intro}
Every isolated population ultimately goes extinct. This happens, even in the absence of adverse environmental variations, because of the discreteness of the individuals and random character of birth and death processes. Extinction risk is a major
negative factor in viability of small populations \cite{Bartlett,Beissinger},
whereas extinction of diseases \cite{Bartlett,Andersson}  is usually beneficial.

Extinction of a \textit{large} population because of the intrinsic noise demands an unusually large fluctuation: a rare sequence of random events when deaths dominate over births. Evaluating the role of rare large fluctuations in far-from-equilibrium systems is hard, and so population extinction, caused by intrinsic noise and environmental variations, has attracted much interest from physicists \cite{Oppenheim,Lindenberg,EK1,Doering,AM1,Kessler,AM2,AM3,Dykman2,KM,AKM1,KMShkl,AKM2,DK,Parker,MS_SIR,AM2010,KMS,KDM,AMS}.
With a few exceptions \cite{Lindenberg,EK1}, these studies assumed well-mixed populations, when spatial degrees of freedom are irrelevant. It has been known, however, since the classical paper of Skellam \cite{Skellam}, that migration of individuals plays a crucial role in a host of natural environments of interest to population biology and epidemiology \cite{Murray}, and in other applications.
An important step forward in quantifying the extinction risk of spatially distributed populations was made by Elgart and Kamenev \cite{EK1}. They considered a population on a discrete lattice that models a refuge of a large but finite size. The population undergoes on-site birth-death processes and migration of individuals between neighboring sites.  Beyond the refuge the conditions are so harsh that they can be modeled by an infinite death rate.  Elgart and Kamenev transformed the master equation for the evolution of a multi-variate probability distribution of the population size into an effective continuous classical mechanics by applying a time-dependent WKB (Wentzel–-Kramers–-Brillouin) approximation that uses the typical on-site population size $K$ in the long-lived state of the population as a large parameter. The time-dependent WKB method yields a Hamiltonian functional and the corresponding Hamilton's equations -- partial differential equations for an effective momentum $p$ (coming from the probability generating function) and a conjugate coordinate $q$ (that, in the deterministic limit, coincides with the populations size). Both $p$, and $q$ depend on the continuous spatial coordinates $\mathbf{x}$ and time $t$. The extinction rate is determined by the classical action calculated along a special trajectory in the (infinite-dimensional) phase space $q(\mathbf{x})$, $p(\mathbf{x})$ of the system \cite{EK1}.

The present paper also deals with extinction of spatially-distributed populations caused by intrinsic noise. We suggest an approach that is closely related to that of Elgart and Kamenev \cite{EK1}, but also differs from it in a number of ways.  First, in addition to scenario I of extinction, considered already in Ref.~\cite{EK1}, we also address scenario II. In scenario I
the zero population size is a repelling fixed point of the on-site deterministic dynamics. In scenario II it is an attracting fixed point, corresponding to what is known in ecology as Allee effect \cite{Allee}. The results in these two extinction scenarios turn out to be quite different. Second, we derive the WKB equations systematically from the master equation for the multi-variate probability distribution. This derivation shows  that a continuous description in space is only valid when the migration rate between the neighboring sites greatly exceeds the on-site process rates. Third, by focusing on the long-lived quasi-stationary distribution of the population size, we formulate a \textit{stationary} WKB theory in terms of the population size (treated as a ``coordinate") and its conjugate momentum. Fourth, an important attribute of this WKB theory is spatial boundary conditions for WKB momentum $p(\mathbf{x},t)$. We derive these boundary conditions, thus correcting an omission in Ref.~\cite{EK1}. Fifth, using the WKB theory, we establish important general properties of the most probable path of the population to extinction.  We show that, in scenario I, the mean time to extinction (MTE) is determined by a  heteroclinic trajectory between two fixed points in the (infinite-dimensional) functional phase space of the system. The first fixed point corresponds to the long-lived quasi-stationary distribution of the population size. The second one corresponds to a zero-population-size state with a nontrivial momentum profile. In scenario II we only have results in the limit of a very strong Allee effect: close to a characteristic bifurcation of the system.  Here again we obtain the solution of the problem
in terms of a heteroclinic connection:  between the fixed point, corresponding to the long-lived quasi-stationary distribution, and a fixed point describing the ``critical nucleus".  In this limit the population extinction problem turns out to be completely integrable, similarly to the integrability of the problem of population explosion close to the saddle-node bifurcation \cite{EK1}.  We explain this integrability by establishing a direct connection between this problem and the over-damped limit of theory of homogeneous nucleation due to Langer \cite{Langer}.  We consider different types of refuge, determined by the conditions at the refuge boundaries and illustrate our results by solving,  analytically and numerically,  three particular population models. In most of this paper we deal with refuges whose spatial sizes are \textit{not} exponentially large in parameter $K \gg 1$. An exception is section \ref{caseB} B where extinction of populations residing in very large refuges is considered (again, for a very strong Allee effect). Surprisingly, we find here an exponentially large reduction in the MTE and predict an optimal refuge size that maximizes the MTE.

The remainder of the paper is organized as follows. Section~\ref{deterministic} includes  important preliminaries which are used in the subsequent sections. It gives an overview of deterministic theory of population dynamics in a refuge: with and without Allee effect, and for different spatial boundary conditions. It also discusses, on a  qualitative level, how the noise-driven population extinction is expected to occur in different cases. Section~\ref{master} presents a stochastic theory of the population dynamics in a refuge. Here we introduce the master equation, focus on the quasi-stationary multi-variate distribution of the population sizes and on the MTE, and formulate a WKB theory aimed at evaluating these quantities. Sections \ref{caseA} and \ref{caseB} analyze population extinction in scenarios I and II, respectively. Here we consider two specific birth-death models in the region of parameters close to their characteristic bifurcations. In this way
we achieve some generality, as the reduced equations, in each of the two cases,  describe a broad class of population models.  We also revisit, in section \ref{nonuniversal}, an additional model problem, exhibiting extinction scenario I. Extinction of populations residing in exponentially large refuges is considered, for a very strong Allee effect,  in section \ref{caseB}. The results are discussed, along with some possible generalizations and unresolved problems, in section~\ref{discussion}.

\section{Deterministic equations and population extinction scenarios}
\label{deterministic}
\subsection{General}
Consider a single population residing in a refuge by which we mean a one-dimensional lattice of $N\gg 1$ sites (or habitat patches) labeled by index $i=1,2,\dots, N$. The population size $n_i$ at each site  varies in time as a result of two types of Markov processes. The first set of processes involves a local, on-site stochastic dynamics of birth-death type, with
birth and death rates $\lambda(n_i)$ and $\mu(n_i)$, respectively, where $\mu(0)=0$. As there is no creation of new individuals ``from vacuum", one has $\lambda(0)=0$.
The second process is random and independent migration of each individual between neighboring sites with migration rate coefficient $D_0$. What happens at the edges of the refuge, $i=1$ and $i=N$, needs to be specified separately; we will deal with this issue a bit later.

Assuming $n_i\gg 1$, one can attempt to neglect fluctuations and describe the population dynamics by deterministic rate equations
\begin{equation}\label{rateeq1}
   \dot{n}_i=\lambda(n_i) -  \mu(n_i)+ D_0 (n_{i-1}+n_{i+1}-2 n_i)\,.
\end{equation}
Established populations are described, in the deterministic limit, by stable steady-state solutions $n_i$ of this set of $N$ coupled equations. According to Eq.~(\ref{rateeq1}), an established population would persist forever. The stochastic picture is markedly different. An unusual sequence of births and (predominantly) deaths can bring the population to the absorbing state $(n_1=0,n_2=0, \dots,n_N=0)$ corresponding to extinction occurring everywhere. This ultimately happens with probability one.

Before dealing with the stochastic problem, however, let us dwell some more on  deterministic rate equations~(\ref{rateeq1}) and their predictions. Let the characteristic population size on a single site, predicted by a steady-state deterministic solution scales as $K\gg 1$.  This implies \cite{Doering,AM2010} that, in the leading order of $K$, one can represent the birth and death rates as
\begin{equation}\label{rates}
    \lambda(n_i)=\mu_0 K \bar{\lambda}(q_i)\;\;\;\;\;\mbox{and} \;\;\;\;\;\mu(n_i)=\mu_0 K \bar{\mu}(q_i),
\end{equation}
where $q_i=n_i/K$ is the rescaled population size at site $i$, $\bar{\lambda}(q_i)\sim \bar{\mu}(q_i) \sim 1$, and $\mu_0$ is a characteristic rate coefficient. Now Eq.~(\ref{rateeq1}) can be rewritten as
\begin{equation}\label{rateeq2}
   \dot{q}_i=\mu_0 f(q_i)+ D_0 (q_{i-1}+q_{i+1}-2 q_i)\,,
\end{equation}
where $f(q_i)=\bar{\lambda}(q_i) - \bar{\mu}(q_i)$ is the rescaled birth-death rate function.

With no migration, $D_0=0$, the on-site deterministic dynamics is determined by the equation $\dot{q}=\mu_0 f(q)$. One fixed point of this equation is  $q=0$, and there are two major cases determined by the sign of derivative $f^{\prime}(q)$ at $q=0$. For $f^{\prime}(0)>0$ (scenario I) the fixed point $q=0$ is
repelling, and the on-site population size, in the absence of migration, flows to an attracting fixed point $q=q_1>0$ that describes an established population. One example of scenario I is the well known SIS model of epidemiology \cite{SIS} for which $\lambda(n)=\lambda_0 n (K-n)$ and $\mu(n)=\mu_0 n$. Here $\bar{\lambda}(q)=R_0 q (1-q)$, $\bar{\mu}(q)=q$, and $f(q)= q (R_0-1-R_0q)$, where $R_0=\lambda_0 K/\mu_0$ is the basic reproduction number. At $R_0>1$  $q=0$ is a repelling point of equation $\dot{q}=\mu_0 f(q)$, whereas $q=q_1=1-1/R_0$ is an attracting point.

In scenario II one has $f^{\prime}(0)<0$. Here fixed point $q=0$ is attracting, and the population gets established, at another attracting fixed point $q=q_2$,  only if the initial population size exceeds a threshold: a repelling fixed point
$q_1$ such that $0<q_1<q_2$. Scenario II accounts, in a simplified way, for a host of \textit{Allee effects} \cite{Allee}. As an example of scenario II we will consider the following three reactions: $A\to 0$, $2A \to 3A$ and $3A \to 2A$ with rate coefficients
$\mu_0$, $\lambda_0$ and $\sigma_0$, respectively \cite{AM2010,EK2}.
Here
$\bar{\lambda}(q)=2q^2/\gamma$ and $\bar{\mu}(q)=q(1+q^2/\gamma)$, where $K=3\lambda_0/(2\sigma_0)$ and $\gamma=8\mu_0\sigma_0/(3\lambda_0^2)$.
At $\delta^2 \equiv 1-\gamma>0$  the system exhibits bistability. Here the zeros of the rescaled birth-death rate function
\begin{equation}\label{fB}
    f(q)=-\frac{1}{\gamma}\,q\,(q-q_1) (q-q_2)
\end{equation}
describe two attracting fixed points, $0$ and $q_2=1+\delta$, and a repelling fixed point $q_1=1-\delta$ such that $0<q_1<q_2$.

Now let us reintroduce deterministic migration and assume that it is much faster than the on-site population dynamics:  $D_0\gg \mu_0$ (the criterion can become less restrictive close to characteristic bifurcations of the on-site population models, see sections \ref{universal} and \ref{caseB}). In this case one can use a continuous spatial coordinate $x$ instead of the discrete index $i$ and replace the discrete Laplacian in Eq.~(\ref{rateeq2}) by the continuous one. This brings about reaction-diffusion equation
\begin{equation}\label{rateeq3}
   \partial_t q=\mu_0 f(q)+ D \partial_x^2 q\,,
\end{equation}
where $D=D_0 h^2$ is the diffusion constant, and $h$ is the lattice spacing. The system size becomes
$L=N h$. Equation~(\ref{rateeq3}), which has been the subject of numerous studies \cite{Murray,Mikhailov}, should be supplemented by spatial boundary conditions. We will separately consider periodic, $q(x+L)=q(x)$, and zero, $q(0)=q(L)=0$, boundary conditions. In the discrete version of the problem, the zero boundary conditions correspond, up to small corrections (see Appendix A), to absorbing boundaries at sites $i=1$ and $i=N$. The absorbing boundaries model, for example, extremely harsh conditions outside of the refuge \cite{EK1,Skellam}. Results for still another type of boundaries -- reflecting walls at $x=0$ and $x=L$ -- can be easily obtained from the results for periodic boundary conditions.

Spatial profiles of established populations are described, in the deterministic theory, by stable steady-state solutions $q=q(x)>0$ of Eq.~(\ref{rateeq3}).  They satisfy ordinary differential equation
\begin{equation}\label{steady}
   D q^{\prime\prime}(x)+\mu_0 f(q)=0\,
\end{equation}
subject to the chosen spatial boundary conditions. The first integral of this equation,
\begin{equation}\label{energy}
    \frac{D}{2\mu_0}\left(q^{\prime}\right)^2+V(q)=const\,,
\end{equation}
with effective potential $V(q)=\int_0^q f(\xi) \,d \xi$, makes the problem soluble in quadratures and yields a phase portrait of the steady states on the plane $(q,q^{\prime})$.
Notably, reaction-diffusion Eq.~(\ref{rateeq3}) is a gradient flow, $\partial_t q =- \delta {\cal F}/\delta q$, where
\begin{equation}\label{freeenergy}
    {\cal F}[q(x,t)]=\int_0^L \,dx\,\left[-\mu_0 V(q)+(1/2) \,D(\partial_x q)^2\right].
\end{equation}
Therefore, it describes a deterministic flow towards a minimum of the
Ginzburg-Landau free energy ${\cal F}[q]$. This property helps identify linearly stable and unstable $x$-dependent solutions, as they correspond to local minima and maxima of ${\cal F}[q]$, respectively \cite{Mikhailov}. Furthermore, it yields a simple
selection rule in cases when, at fixed $L$, there are multiple solutions of Eq.~(\ref{steady}) with periodic boundary conditions: the solution with the maximum spatial period (equal to $L$) is selected when starting from a generic initial condition \cite{periodic}.

\begin{figure}[ht]
\includegraphics[width=2.0 in,clip=]{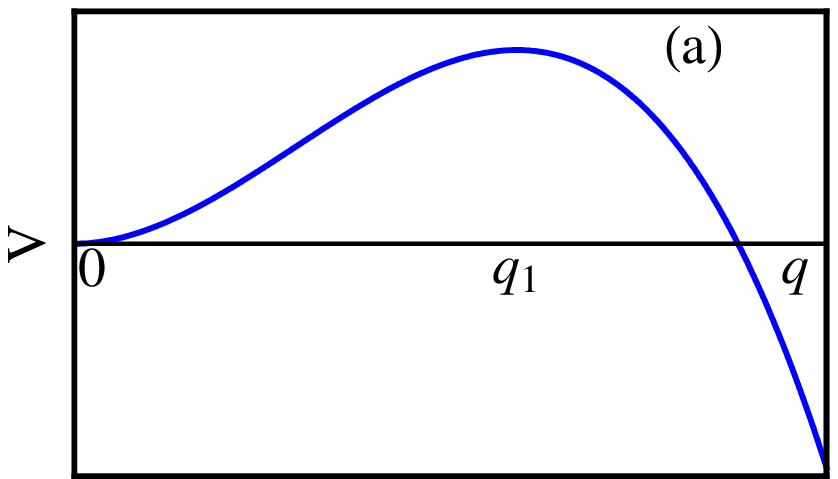}
\includegraphics[width=2.0 in,clip=]{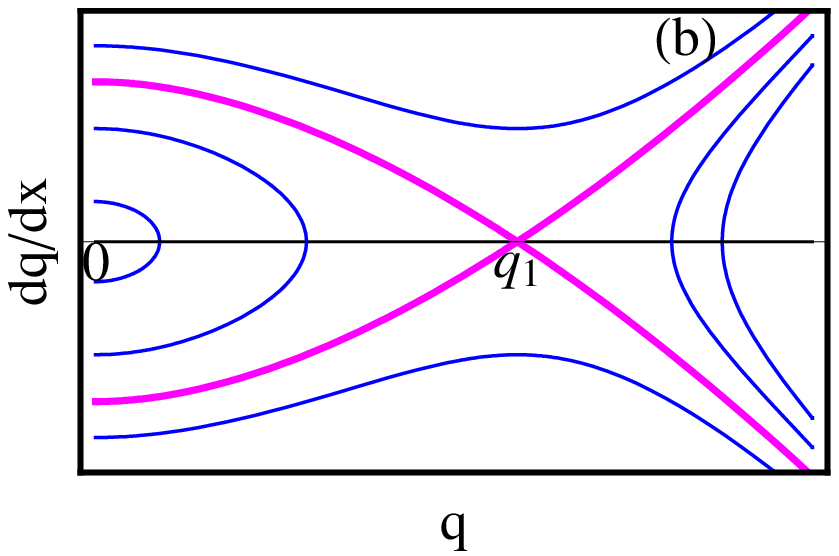}
\caption{(color online) Effective potential $V(q)$ and phase portrait $(q,q^{\prime})$ for steady-state solutions of Eq.~(\ref{rateeq3}) in scenario I (no Allee effect).}
\label{A1}
\end{figure}

\begin{figure}[ht]
\includegraphics[width=2.0 in,clip=]{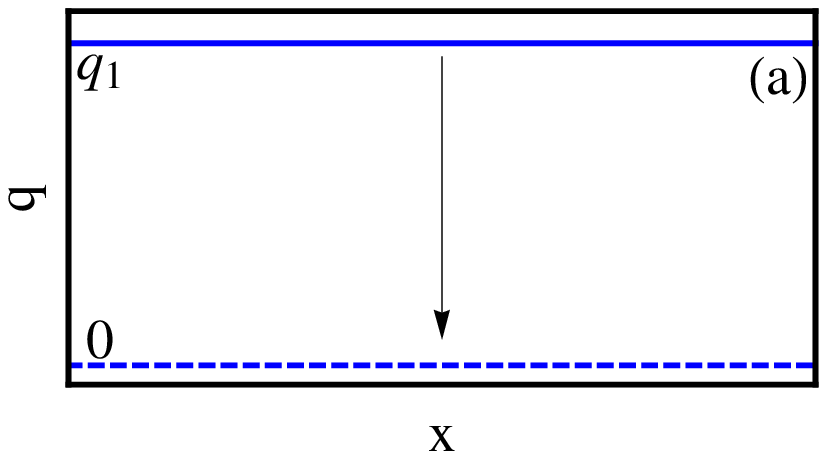}
\includegraphics[width=2.0 in,clip=]{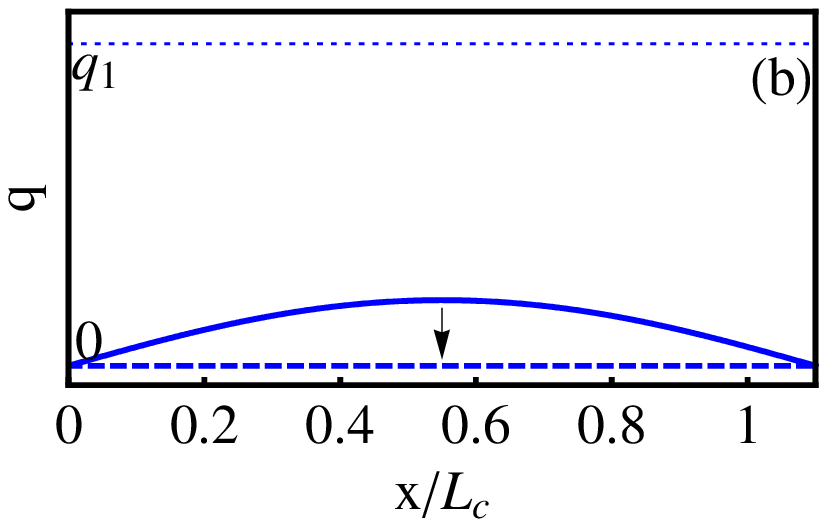}
\includegraphics[width=2.0 in,clip=]{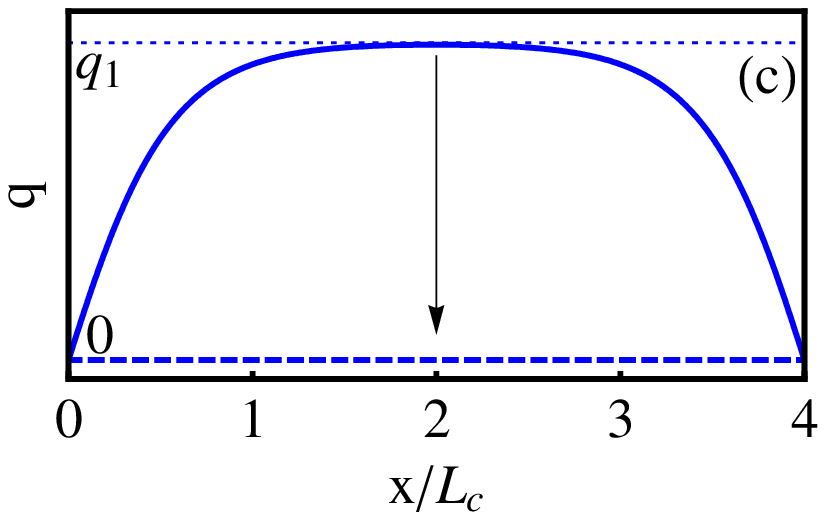}
\caption{(color online) Steady-state solutions of Eq.~(\ref{rateeq3}) in scenario I (no Allee effect) for periodic (a) and zero (b and c) boundary conditions in space. The arrows indicate extinction transitions driven by rare large fluctuations. $L=1.1 L_c$ (b) and $4 L_c$ (c). $R_0=2$, so $L_c=\pi(D/\mu_0)^{1/2}$,  see Eq.~(\ref{Lc}).}
\label{A2}
\end{figure}

\subsection{Scenario I}

What is the steady state in scenario I, as exemplified by the spatio-temporal SIS model? Figure~\ref{A1} shows effective potential $V(q)= (R_0-1) q^2/2-R_0q^3/3$ and the resulting phase portrait $(q,q^{\prime})$
at $R_0>1$. The only non-trivial steady-state solution, obeying periodic boundary conditions, is the $x$-independent solution $q=q_1$, depicted in Fig.  \ref{A2}a.  Introducing intrinsic noise, we will see that the most probable path of this population to extinction is such
that the population size drops to zero uniformly on the whole interval $0\leq x\leq L$. For large systems, the MTE is very long in this case, being exponentially large in $K L/h=K N$.

For the zero boundary conditions, an $x$-dependent steady state corresponds to a phase trajectory inside the separatrix in Fig. \ref{A1}. Such steady states, depicted in Figs.~\ref{A2} b and c,  exist only if the system size $L$ is larger than critical size
\begin{equation}\label{Lc}
    L_c=\pi \sqrt{\frac{D}{\mu_0(R_0-1)}}\,.
\end{equation}
This quantity can be obtained from Eq. (\ref{steady}) linearized around $q=0$. At $L<L_c$ there is only trivial solution: no established population. The $x$-dependent solution emerges, at $L=L_c$, via a transcritical bifurcation. At $L\gg L_c$ the population size is close to $q_1$ everywhere except in boundary layers, with thickness of order of $L_c$, at $x=0$ and $x=L$.  At $L>L_c$ the most probable path to noise-driven extinction for the zero boundary conditions is such
that the population size drops to zero uniformly on the whole interval $0\leq x\leq L$. As we will see in section \ref{caseA} (see also Ref. \cite{EK1}), for $L\gg L_c$
the MTE is again exponentially long in parameter $KN$. It becomes much shorter as $L$ approaches $L_c$, see section \ref{caseA}.

\subsection{Scenario II: Allee effect}

Now consider scenario II, on the example of three reactions $A\to 0$ and $2A \rightleftarrows 3A$. At $0<\gamma<1$, that is $0<\delta<1$, the effective potential,
\begin{equation}\label{VB}
     V(q)=-\frac{q^2}{2}+\frac{2 q^3}{3 \gamma}-\frac{q^4}{4 \gamma},
\end{equation}
has two maxima: at $q=0$ and $q=q_2$, and which of the steady-state solutions $q=0$ and $q=q_2$ ``wins"  depends on which of the maxima is higher \cite{Murray,Mikhailov}.

\subsubsection{Strong Allee effect}

\begin{figure}[ht]
\includegraphics[width=2.0 in,clip=]{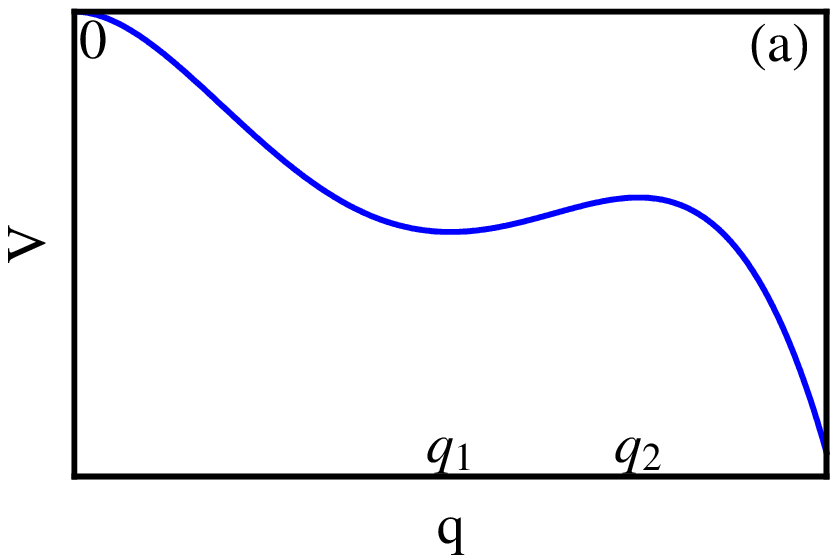}
\includegraphics[width=2.0 in,clip=]{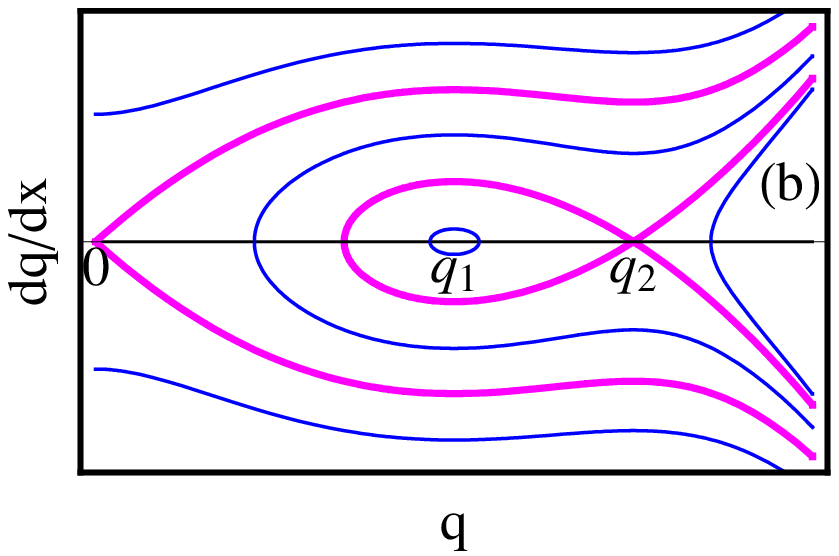}
\caption{(color online) Effective potential $V(q)$ and phase portrait $(q,q^{\prime})$ for steady-state solutions of Eq.~(\ref{rateeq3}) for a strong Allee effect, $V(0)>V(q_2)$.}
\label{Bstrong1}
\end{figure}

\begin{figure}[ht]
\includegraphics[width=2.0 in,clip=]{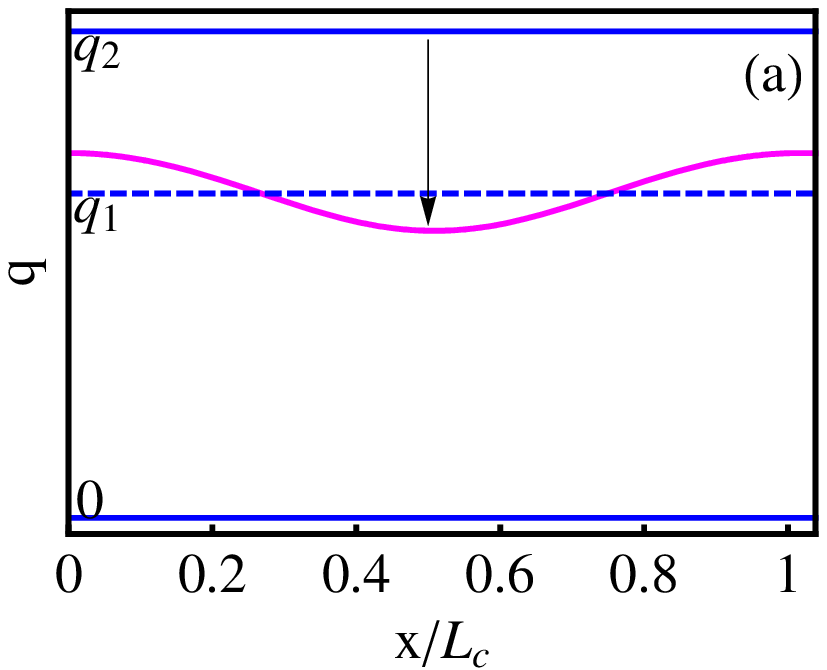}
\includegraphics[width=2.0 in,clip=]{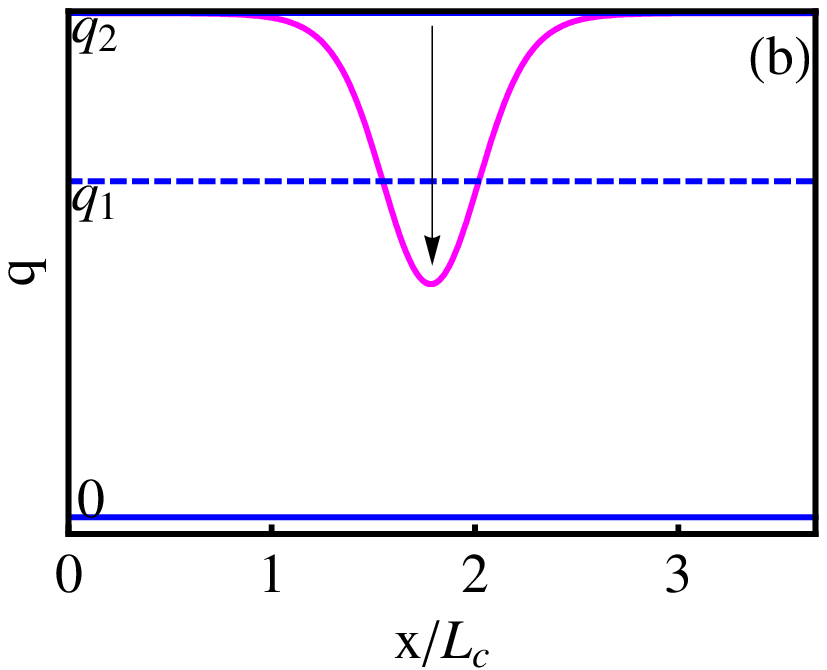}
\caption{(color online) Linearly stable states $q=q_2$ and $q=0$, linearly unstable state $q=q_1$ and critical nucleus $q_c(x)$ for a strong Allee effect and periodic boundary conditions. The system size $L=1.04 L_c$ (a) and $3.7 L_c$ (b), where $L_c$ is defined in Eq.~(\ref{Lc1}). The arrows indicate  transitions, driven by rare large fluctuations and leading to a rapid extinction. Parameter $\gamma=24/25$, so $\delta=1/5$, and $L_c=2\pi(3D/\mu_0)^{1/2}$.}
\label{Bstrong2}
\end{figure}

Figures \ref{Bstrong1} and \ref{Bstrong2} illustrate the case of $V(q_2)<V(0)$: a strong Allee effect. In our example this occurs at $8/9<\gamma<1$, or $0<\delta<1/3$. For periodic boundary conditions, the only linearly stable nontrivial steady-state solution is the $x$-independent solution $q=q_2$. A sufficiently large perturbation, however, triggers a \textit{deterministic} transition from $q=q_2$ to the trivial solution $q=0$
that is also linearly stable. An important attribute of this metastability is presence of the ``critical nucleus": an $x$-dependent solution $q_c(x)$ of Eq.~(\ref{steady}) that is linearly unstable under the dynamics of Eq.~(\ref{rateeq3}). A small perturbation around the critical nucleus brings the system either to $q=0$ or to $q=q_2$. The critical nucleus  is selected by the system size $L$ and corresponds to a phase trajectory inside the internal separatrix shown in Fig. \ref{Bstrong1}b. The critical nucleus exists only for $L>L_c$, where
\begin{equation}\label{Lc1}
    L_c=\pi \sqrt{\frac{2 D (1+\delta)}{\mu_0 \delta}}\,,
\end{equation}
as can be obtained from Eq. (\ref{steady}), linearized around $q=q_1$, with periodic boundary conditions.  At $L\gg L_c$ the critical nucleus coincides with the internal separatrix in Fig. \ref{Bstrong1}b. For  $f(q)$ from Eq.~(\ref{fB}) (a cubic polynomial), the critical nucleus can be found analytically, in terms of elliptic functions, by integrating the first-order equation~(\ref{energy}) and choosing the solution $q(x)$ with period equal to the system size $L$.  A more practical
way is to solve Eq.~(\ref{steady}) numerically, by shooting. One solves numerically an \textit{initial-value} problem for Eq.~(\ref{steady}) starting, at $x=0$, from some $q(0)\in(q_1,q_2)$ and $q^{\prime}(0)=0$. Parameter $q(0)$ is varied until the numerical solution exhibits a single full-period oscillation, so that $q(L)\simeq q(0)$ and $q^{\prime}(L)\simeq 0$. Figure~\ref{Bstrong2} shows the critical nuclei for two different values of $L>L_c$. Note that a critical nucleus corresponds to a local \textit{maximum}
of free energy (\ref{freenergy}) \cite{Mikhailov}.

The presence of a critical nucleus in the deterministic theory plays a pivotal role in the noise-driven extinction of an established population exhibiting a strong Allee effect. Indeed, a large fluctuation of the size of stochastic population residing around $q=q_2$ can create critical nucleus $q_c(x)$. The further population dynamics toward extinction proceeds ``downhill", that is essentially deterministically. What happens at $L\gg L_c$, see Fig.~\ref{Bstrong2}b, is intuitively clear, and will be supported by our quantitative results in section \ref{caseB}. Here the rate of noise-induced creation of the critical nucleus is exponentially small in $K$ but independent of $L$  (unless $L$ is exponentially large in $K$).  Once having passed the critical nucleus, the solution $q(x,t)$ of Eq.~(\ref{rateeq3}) develops, on a time scale $\sim \mu_0^{-1}$, two outgoing deterministic ``extinction fronts". In our example of three reactions the deterministic front solution can be found analytically \cite{Murray,Mikhailov}. The extinction fronts propagate with speed
\begin{equation}\label{speed}
    c=\sqrt{\frac{\mu_0D}{2(1-\delta^2)}}\,\left(1-3 \delta\right)
\end{equation}
and drive the whole population to extinction on a time scale $\sim L/(\mu_0 D)^{1/2}$. Therefore, unless the system size $L$ is exponentially large in $K$, it is the creation of a single critical nucleus that serves as the extinction bottleneck. That is, the MTE is determined here by the mean creation time of the critical nucleus. This quantity does not include an exponential dependence on the system size $L$ and is therefore much shorter than in scenario I. Now, what happens when $L$ is above $L_c$ but close to it? We will show that here too the most probable path to extinction corresponds to a large fluctuation bringing the population from $q=q_2$ to critical nucleus $q_c(x)$, see Fig.~\ref{Bstrong2}a, and \textit{not} to the $x$-independent unstable state $q=q_1$.

For a strong Allee effect and zero boundary conditions, there is only (linearly stable) trivial steady state $q=0$: no established population.

\begin{figure}[ht]
\includegraphics[width=2.0 in,clip=]{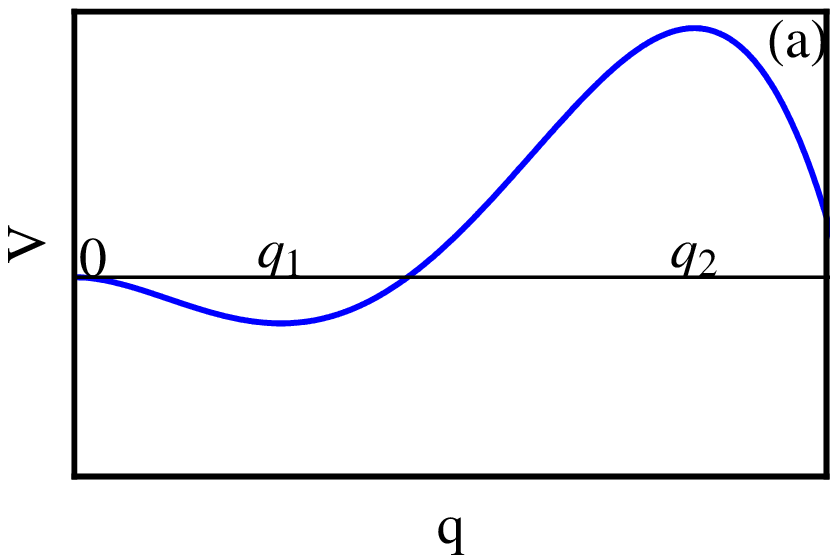}
\includegraphics[width=2.0 in,clip=]{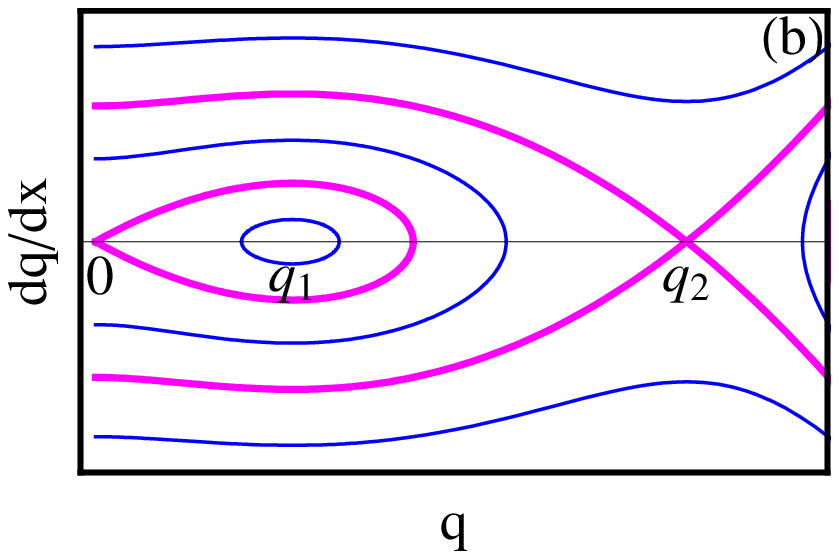}
\caption{(color online) Effective potential $V(q)$ and phase portrait $(q,q^{\prime})$ for steady-state solutions of Eq.~(\ref{rateeq3}) for a weak Allee effect, $V(0)<V(q_2)$.}
\label{Bweak1}
\end{figure}

\begin{figure}[ht]
\includegraphics[width=2.0 in,clip=]{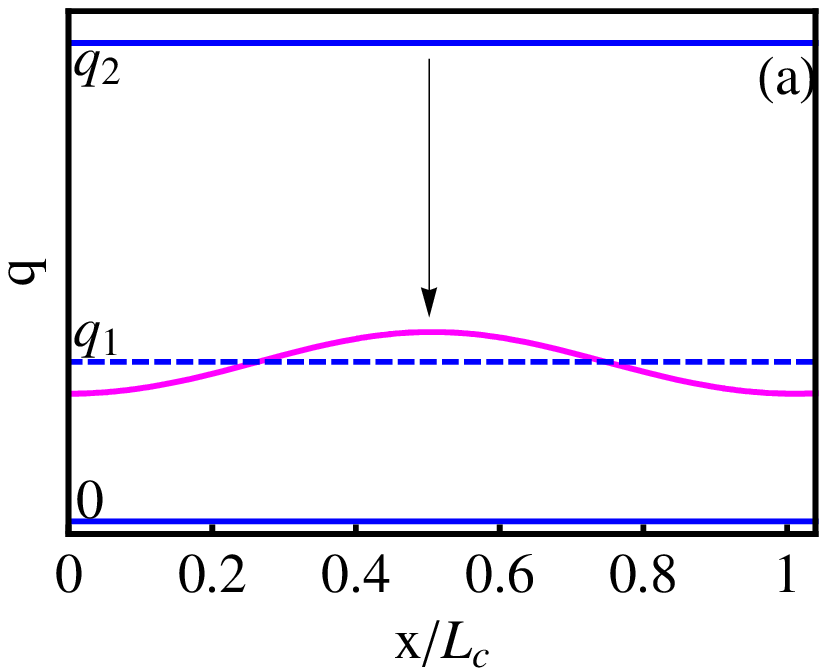}
\includegraphics[width=2.0 in,clip=]{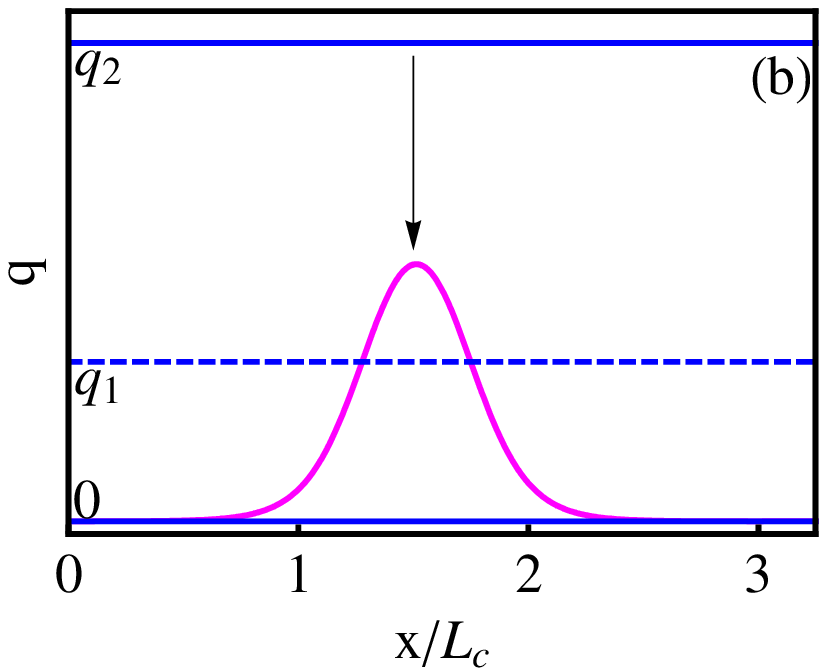}
\caption{(color online) Linearly stable steady states $q=q_2$ and $q=0$, linearly unstable state $q=q_1$ and critical nucleus $q=q_c(x)$ for a weak Allee effect and periodic boundary conditions. The system size $L=1.4 L_c$ (a) and $3.25 L_c$ (b), where $L_c$ is defined in Eq.~(\ref{Lc1}). The arrows indicate extinction transitions driven by rare large fluctuations. $\gamma=3/4$, so $\delta=1/2$ and $L_c=\pi(6D/\mu_0)^{1/2}$.}
\label{Bweak2}
\end{figure}

\subsubsection{Weak Allee effect}

For a weak Allee effect one has  $V(q_2)>V(0)$, as illustrated in
Figs.~\ref{Bweak1} and \ref{Bweak2}. In our example of three reactions this case corresponds to $0<\gamma<8/9$, or $1/3<\delta<1$. For periodic boundary conditions there are two linearly stable $x$-independent steady states, $q=q_2$ and $q=0$, and the linearly unstable $x$-independent state $q=q_1$. There is also critical nucleus $q=q_c(x)$, described by a phase trajectory located inside the internal separatrix in Fig.~\ref{Bweak1}b; it is selected by the system size $L$. The critical nucleus exists when $L>L_c$, where $L_c$ is given by Eq.~(\ref{Lc1}). At $L\gg L_c$ the critical nucleus is described by the internal separatrix  of Fig. \ref{Bweak1}b. Here the population size, corresponding to the critical nucleus $q_c(x)$, is close to  zero everywhere except in a narrow region with thickness $\sim L_c$. 
What is the most probable path of the population toward noise-driven extinction?
Here one has to choose between two paths. In the first path the population size goes down from $q=q_2$ to the $x$-independent unstable state $q=q_1$ on the whole interval $0<x<L$ and then continues to fall, almost deterministically, to zero. In the second path the population size goes down from $q=q_2$ to the critical nucleus and then, almost deterministically, to zero.
For $L\gg L_c$ the MTE  involves, for each of the two options, an exponential dependence on $L$, so it can be very long.

\begin{figure}[ht]
\includegraphics[width=2.0 in,clip=]{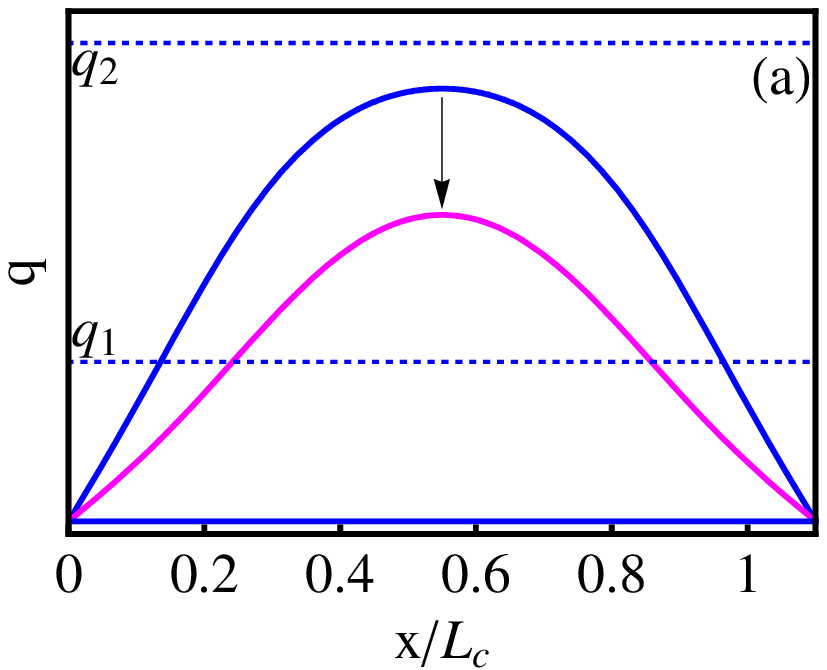}
\includegraphics[width=2.0 in,clip=]{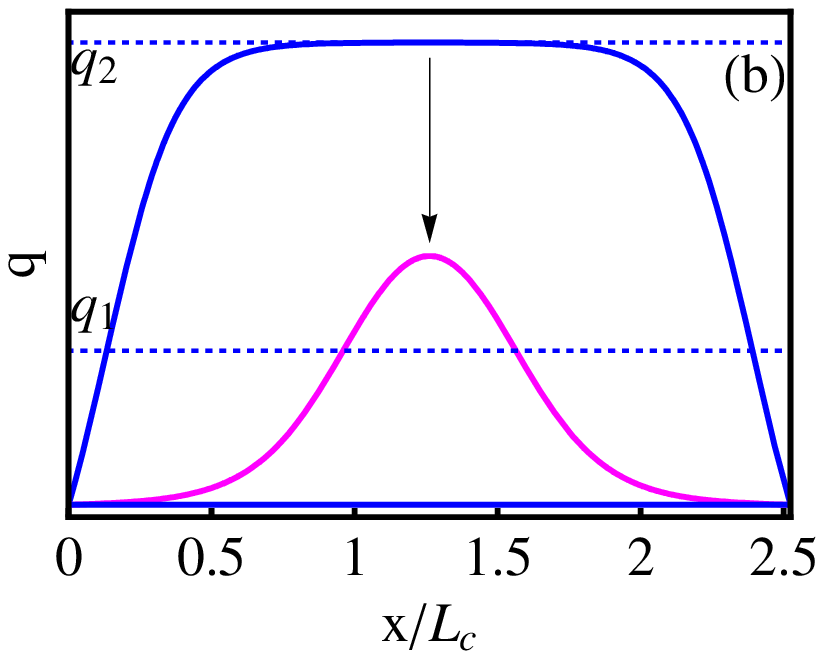}
\caption{(color online) The $x$-dependent linearly stable steady state (the upper curve), trivial stable state $q=0$, linearly unstable state $q=q_1$ and critical nucleus $q=q_c(x)$ for a weak Allee effect  and zero boundary conditions for $L=1.1 L_c$ (a) and $2.52$ (b). The arrows indicate the extinction transitions driven by rare large fluctuations. $\gamma=3/4$, so $\delta=1/2$, and $L_c\simeq 6.026 (D/\mu_0)^{1/2}$.}
\label{Bweak3}
\end{figure}

For the zero boundary conditions there are two linearly stable steady states: an $x$-dependent state and the trivial state $q=0$.  There is also critical nucleus: an $x$-dependent unstable steady state. These solutions are depicted in Fig.~\ref{Bweak3} a and b. Each of the $x$-dependent solutions is selected by the system size $L$ and described by a phase trajectory located between the two separatrices in Fig. \ref{Bweak1}b. Among them there is a limiting phase trajectory such that the stable steady state, at given $L$, corresponds to a phase trajectory located between the limiting phase trajectory and the external separatrix. In its turn, the critical nucleus, for the same $L$, corresponds to a phase trajectory that lies between the limiting phase trajectory and the internal separatrix. The $x$-dependent solutions, both stable and unstable, exist when the system size $L$ is larger than a critical size $L_c$ [which is different from $L_c$ given by Eq.~(\ref{Lc1})]. The critical size $L_c$ scales as $(D/\mu_0)^{1/2}$ and also depends on $\delta$. For $L\gg L_c$ the linearly stable steady state corresponds to the external separatrix and is therefore close to $q_2$ everywhere except in boundary layers with thickness $\sim L_c$ at $x=0$ and $x=L$. In its turn, the critical nucleus corresponds, at $L\gg L_c$, to the internal separatrix and therefore coincides with the critical nucleus obtained for periodic boundary conditions. At $L=L_c$ the stable and unstable solutions merge. At $L<L_c$ there is only trivial steady state $q=0$ which is linearly stable. The most probable path to extinction at $L>L_c$ corresponds to a large fluctuation that brings
the population size from the stable state down to the critical nucleus, see  Fig.~\ref{Bweak3} a and b.


\section{Master equation and WKB approximation}
\label{master}

\subsection{Governing equations}
\label{governing}

Now let us return to the discrete-lattice model and describe \textit{stochastic} dynamics of the population. This can be done in terms of evolution of the multivariate probability distribution
$P(\mathbf{n},t)= P(n_1,n_2,\dots,t)= P(\hat{\mathbf{n}},n_i,t)$,
where $i=1,2, \dots, N$, and $\hat{\mathbf{n}}$ denotes the vector of all $n$'s not explicitly written, see \textit{e.g.} Ref. \cite{Gardiner}.
This probability distribution is assumed to be identically zero if any of $n_i$ is negative. For the continuous-time Markov processes of birth, death and migration, the master equation for $P(\mathbf{n},t)$ has the following form:
\begin{widetext}
\begin{eqnarray}
   \partial_t P(\mathbf{n},t) &=&  \sum_{i=1}^N\Big\{\lambda(n_i-1) P(\hat{\mathbf{n}},n_i-1,t)+\mu(n_i+1) P(\hat{\mathbf{n}},n_i+1,t) -[\lambda(n_i)+\mu(n_i)] P(\mathbf{n},t)\Big\}\nonumber \\
 &+&D_0\sum_{i=1}^N \Big\{(n_{i-1}+1) P(\hat{\mathbf{n}},n_{i-1}+1,n_i-1,t)+ (n_{i+1}+1) P(\hat{\mathbf{n}},n_i-1,n_{i+1}+1,t) -2 n_i P(\mathbf{n},t)\Big\}\,.\label{master1}
\end{eqnarray}
\end{widetext}
This equation holds as it is for a periodic lattice with period $N$. For absorbing boundaries the migration terms $i=1$ and $i=N$ are slightly different, see Appendix A.
Of a primary interest for us is the instantaneous extinction rate, or extinction probability flux:
\begin{eqnarray}
   \partial_t P(\mathbf{0},t) &=& \mu(1)[P(1,0,\dots,0,t)+P(0,1,\dots,0,t)+\dots \nonumber \\
    &+&P(0,0,\dots,1,t)]\,.\label{current}
\end{eqnarray}
We will continue to assume that $K\gg 1$. Furthermore, we will assume in most of the paper (except in section \ref{caseB} B) that the system size is not too large: not exponentially large in $K$. In this case, extinction of an established population proceeds, in the probabilistic language,  as follows. During the relatively short \textit{relaxation time} $t_r$, determined by the deterministic rate equation (\ref{rateeq1}), the system approaches a quasi-stationary state, where $P(\mathbf{n},t)$ is sharply peaked at the relevant steady-state solution of Eq.~(\ref{rateeq1}). At $t\gg t_r$ the quasi-stationary probability slowly decays in time,
\begin{equation}\label{qsd1}
    P(\mathbf{n},t) \simeq \pi(\mathbf{n})\,e^{-t/T_e}\,,\;\;\;n_i=0,1,2,\dots\,,
\end{equation}
except for $\mathbf{n}=\mathbf{0}$ that corresponds to a complete extinction. The decay rate $1/T_e$ is the lowest positive eigenvalue of the time-dependent master equation (\ref{master1}). This eigenvalue is special: it turns out to be exponentially small with respect to $K\gg 1$ \cite{toosmallD}.  The probability of complete extinction $P(\mathbf{0},t)=P(0,0, \dots,0,t)$ slowly grows in time:
\begin{equation}\label{P0}
    P(\mathbf{0},t) \simeq 1-e^{-t/T_e}\,.
\end{equation}
In this regime the decay time $T_e$ is equal to the MTE, whereas the probability distribution of extinction times is an exponential distribution with mean $T_e$, see \textit{e.g.} Ref. \cite{AM2}. Using Eqs.~(\ref{qsd1}) and (\ref{P0}), we can rewrite Eq.~(\ref{master1}) as a linear eigenvalue problem
for the quasi-stationary distribution $\pi(\mathbf{n})$:
\begin{widetext}
\begin{eqnarray}
  && \sum_{i=1}^N\Big\{\lambda(n_i-1) \,\pi(\hat{\mathbf{n}},n_i-1)+\mu(n_i+1) \,\pi(\hat{\mathbf{n}},n_i+1) -[\lambda(n_i)+\mu(n_i)] \,\pi(\mathbf{n})\Big\}\nonumber \\
 &&+ D_0\sum_{i=1}^N \Big\{(n_{i-1}+1) \;\pi(\hat{\mathbf{n}},n_{i-1}+1,n_i-1)+ (n_{i+1}+1) \,\pi(\hat{\mathbf{n}},n_i-1,n_{i+1}+1) -2 n_i \,\pi(\mathbf{n})\Big\}= -\Lambda \pi(\mathbf{n})\,,\label{qsd2}
\end{eqnarray}
\end{widetext}
(except for $\mathbf{n}=\mathbf{0}$) for the lowest positive
eigenvalue $\Lambda= 1/T_e$.  Once $\pi(\mathbf{n})$ is determined, $\Lambda$ can be found from relation
\begin{eqnarray}
   \Lambda &=& \mu(1)[\pi(1,0,\dots,0)+\pi(0,1,\dots,0)+\dots \nonumber \\
    &+&\pi(0,0,\dots,1)] \label{E}
\end{eqnarray}
following from Eqs.~(\ref{current})-(\ref{P0}).

For $K\gg 1$ and $n_i\gg 1$ we can treat $q_i=n_i/K$ as continuous quantities and solve Eq.~(\ref{qsd2}) in WKB approximation which generalizes to spatial populations the stationary WKB method \cite{Kubo,Kessler,Dykman2,DK,AM2010,KDM,AMS,MS,EscuderoKamenev} previously employed for well-mixed populations. The WKB ansatz is
\begin{equation}\label{WKB}
    \pi(\mathbf{n})= \exp\left[-K S(\mathbf{q})\right]\,.
\end{equation}
Our goals are to accurately evaluate the leading-order contribution to $\ln (\mu_0 T_e)$ and to find the most probable path of the population to extinction.  We plug Eqs.~ (\ref{rates}) and (\ref{WKB})
in Eq.~(\ref{master1}) and neglect term $-\Lambda \pi(\mathbf{n})$ which is expected to be exponentially small in $K\gg 1$.  In the leading order in $1/K$ this procedure yields a stationary Hamilton-Jacobi equation
\begin{equation}\label{HJ}
  H (\mathbf{q},\partial_\mathbf{q} S)=0
\end{equation}
with an effective classical Hamiltonian with $N$ degrees of freedom,
\begin{eqnarray}
  &&\!\!\!\!\!\!\!\!H (\mathbf{q},\mathbf{p}) =\mu_0 \sum_{i=1}^{N} \left[\bar{\lambda}(q_i) \left(e^{p_{i}}-1\right) +\bar{\mu}(q_i) \left(e^{-p_{i}}-1\right)\right]\nonumber\\
 && \!\!\!\!\!\!\!\!\!+ D_0 \sum_{i=1}^N \left[q_{i-1}\left(e^{p_i-p_{i-1}}-1\right)+q_{i+1} \left(e^{p_i-p_{i+1}}-1\!\right)\right]\!,\label{H}
\end{eqnarray}
where $p_i=\partial_{q_i} S$. This lattice Hamiltonian, and corresponding Hamilton's equations -- a set of $2N$ ordinary differential equations for $\dot{q}_i(t)$ and $\dot{p}_i(t)$ -- is a proper framework for dealing with population extinction for any relation between the migration rate coefficient $D_0$ and the characteristic rate coefficient $\mu_0$ of the on-site dynamics \cite{toosmallD}.

In the following we will only consider the limit when, as in section \ref{deterministic},  migration between the neighboring sites is much faster than the on-site population dynamics: $D_0\gg \mu_0$ (the criterion becomes softer close to  bifurcations of the on-site models, see sections \ref{universal} and \ref{caseB}). In this regime the quasi-stationary distribution $\pi(\mathbf{n})$ and, as a consequence,  the classical action $S(\mathbf{q})$ are slowly varying functions of $\mathbf{n}$ and $\mathbf{q}$, respectively. This implies that the difference between the momenta $p_i$ on neighboring sites is much smaller than unity. Taylor-expanding the migration term $H_m$ of  Hamiltonian (\ref{H}) (the term proportional to $D_0$) up to second order, we obtain
\begin{eqnarray}
  H_m(\mathbf{q},\mathbf{p}) &=& D_0\sum_{i=1}^N\left[-\left(q_i-q_{i-1}\right)\left(p_i-p_{i-1}\right)\right. \nonumber \\
  &+&\frac{1}{2}(q_{i}+q_{i-1})\left(p_i-p_{i-1}\right)^2\,]\,. \label{H10}
\end{eqnarray}
The slow variation of $q_i$ and $p_i$ with $i$ calls for a continuous description. We introduce a continuous spatial coordinate $x$ instead of the discrete index $i$ and arrive at an effective continuum classical mechanics. The Hamiltonian functional is
\begin{equation}
 H\left[q(x,t),p(x,t)\right] =\frac{1}{h}\int_0^L \,dx\, w \,,\label{H20}
\end{equation}
with density
\begin{equation}
 w=H_0(q,p) - D\left[\partial_x q\, \partial_x p-q \left(\partial_x p\right)^2\right] \label{H21}
\end{equation}
and on-site Hamiltonian
\begin{equation}\label{onsiteH}
    H_0(q,p)=\mu_0\left[\bar{\lambda}(q) \left(e^p-1\right) + \bar{\mu}(q) \left(e^{-p}-1\right)\right].
\end{equation}
Note the presence of two diffusion terms inside the square brackets in Eq.~(\ref{H21}). The first term describes deterministic diffusion,  the second one describes fluctuations of diffusion.  Hamiltonian, related to Eq.~(\ref{H20}) by canonical transformation
$\mathcal{Q}=q e^{-p},\;\mathcal{P}=e^p$, was obtained by Elgart and Kamenev \cite{EK1} who employed the probability generating function in conjunction with a \textit{time-dependent} WKB theory. Note that the two diffusion terms in Eq.~(\ref{H21}) add up to $- D \,\partial_x \mathcal{Q}\, \partial_x \mathcal{P}$ in canonical variables $\mathcal{Q}$ and $\mathcal{P}$. This simplification, and the somewhat simpler form of the on-site Hamiltonian, can be advantageous, see section \ref{nonuniversal}.

The Hamilton's equations of motion,
\begin{eqnarray}
\partial_t q=h\,\frac{\delta H}{\delta p}&=&\mu_0\left[\bar{\lambda}(q)e^p - \bar{\mu}(q)e^{-p}\right]\nonumber \\
&+&D \left[\partial_x^2q -2\partial_x\left(q\partial_x p\right)\right]\,,
\label{p100}\\
\partial_t p=-h\,\frac{\delta H}{\delta q} &=& -\mu_0 \left[\bar{\lambda}^{\prime}(q)(e^p-1) + \bar{\mu}^{\prime}(q)(e^{-p}-1)\right]\nonumber \\
&-&D\left[\partial_x^2 p +\left(\partial_x p\right)^2\right]\,,
\label{p110}
\end{eqnarray}
are partial differential equations for continuous variables $q(x,t)$ and $p(x,t)=h \,\delta S/\delta q$ \cite{positive}. Note that, for the purpose of solving stationary Hamilton-Jacobi equation  (\ref{HJ}), time as appears in Hamilton's Eqs.~(\ref{p100}) and (\ref{p110}) is merely a way of parametrizing phase space trajectories. It is not necessarily related to the original time entering Eqs.~(\ref{master1})-(\ref{P0}) for the evolution of probabilities. To remind the reader, Eqs.~(\ref{H}) and~(\ref{H10}), as well as Eqs.~(\ref{master1}) and ~(\ref{qsd2}), are only valid for periodic systems;  absorbing boundaries are considered in Appendix A. It is important, however, that continuous equations (\ref{H20})-(\ref{p110}) are valid in the case of absorbing boundaries as well.

For all types of spatial boundaries, continuous Eqs.~(\ref{p100}) and (\ref{p110}) must be complemented with spatial boundary conditions. This circumstance was left unattended in Ref.~\cite{EK1}. For periodic systems the spatial boundary conditions are or course $q(0,t)=q(L,t)$  and $p(0,t)=p(L,t)$. For reflecting boundaries they are also straightforward: $\partial_x q(0,t)=\partial_x q(L,t)=0$ and $\partial_x p(0,t)=\partial_x p(L,t)=0$. The case of absorbing boundaries is a bit more involved, and we derive the corresponding boundary conditions in Appendix A.  Up to small corrections ${\cal O}(\mu_0/D_0)^{1/2}\ll 1$, they turn out to be zero conditions both for the coordinate, and for the momentum:
$q(0,t)=q(L,t)=0$, and $p(0,t)=p(L,t)=0$.

\subsection{Activation trajectories}
\label{hetero}

Now let us return to Eqs.~(\ref{WKB}) and (\ref{HJ}) that describe, in WKB approximation, the quasi-stationary distribution $\pi(\mathbf{n})$. This distribution is smooth and has its (Gaussian) maximum at $q(x)=q_s(x)$. Therefore, in order to find $\pi(\mathbf{n})$, one
needs to find a particular solution of Hamilton-Jacobi Eq.~(\ref{HJ}) such that its variational derivative vanishes at $q=q_s(x)$: $\left.\delta S/\delta q\right|_{q_s(x)}=0$.
Setting $S\left\{q_s(x)\right\}=0$, we define
$S\left\{q(x)\right\}$ uniquely as the solution of Eq.~(\ref{HJ}).
Once this solution is known, one can use Eq.~(\ref{E}) to evaluate the MTE up to pre-exponential factors:
\begin{equation}\label{rev}
    \ln (\mu_0 T_e) \simeq K S(\mathbf{0})\,.
\end{equation}
In order to calculate $S(\mathbf{0})$ we will use Hamilton's equations (\ref{p100}) and (\ref{p110})
that describe trajectories in the functional phase space $\left\{q(x),p(x)\right\}$.
As Hamiltonian~(\ref{H20}) does not depend explicitly on time, it is a constant of motion. Furthermore, as Hamilton-Jacobi equation (\ref{HJ}) is stationary, we should only consider trajectories,  for which this constant of motion -- the total energy of the effective mechanical system -- is zero. The simplest among zero-energy \textit{trajectories} are deterministic, or \textit{relaxation}, trajectories: solutions of Eqs.~(\ref{p100}) and (\ref{p110}) with $p(x,t)=0$. Here Eq.~(\ref{p100}) reduces to the deterministic reaction-diffusion equation~(\ref{rateeq3}),
whereas Eq.~(\ref{p110}) is satisfied trivially.

The quasi-stationary distribution $\pi(\mathbf{n})$ is peaked at what we call fixed point A: (functional) fixed point $q(x)=q_s(x),\, p(x)=0$ of Eqs.~(\ref{p100}) and (\ref{p110}). Therefore, the phase trajectory we are interested in for the purpose of calculating $S(\mathbf{0})$ should start, at $t=-\infty$, at fixed point $A$.  In both extinction scenarios I and II there are a stable manifold $p(x)=0$, and an unstable
manifold $p(x)\neq0$, emanating from fixed point A, see Appendix B. In the discrete lattice formulation, each of these two manifolds is $N$-dimensional  and is embedded into zero-energy hyper-surface $H\{q(x),p(x)\}=0$.

For any phase trajectory that originates from fixed point $A$ at $t=-\infty$, we can write
$$
 S\{q(x,T)\}=\frac{1}{h}\int_{-\infty}^T dt\, \int_0^L p(x,t)\,\partial_t q(x,t)\, dx \,.
$$
In view of Eq.~(\ref{rev}), we only need to consider phase trajectories that reach extinction hyper-plane $q(x)=0$ [so that $q(x,t)$ vanishes at all $x$]. Relaxation trajectories, $p=0$, that exit fixed point A, cannot reach the extinction hyper-plane, so we need an \textit{activation} trajectory, $p\neq 0$, for this purpose. For extinction scenario I, a crucial property of the activation trajectory can be established under quite general assumptions.  The activation trajectory must approach, at $t=+\infty$, another fixed point which we call fixed point B. It involves $q(x)=0$, see Fig.~\ref{A2}, and $p(x)=p_e(x)$: the non-trivial steady-state solution of Eq.~(\ref{p110}) with $q(x)=0$ and proper spatial boundary conditions.
That is, in scenario I the activation trajectory must be a heteroclinic connection AB [or instanton, see Ref. \cite{instantons} for a review on instantons] in functional phase space $\left\{q(x),p(x)\right\}$.
The proof of this statement is presented in Appendix B2; it relies on the structure of the phase space of Eqs.~(\ref{p100}) and (\ref{p110}) and, in particular, on the presence
and linear stability properties of (zero-energy) fixed points of Eqs.~(\ref{p100}) and (\ref{p110}).

For extinction scenario II the structure of the phase space is more complicated, and we cannot make an equally general statement about the properties of the activation trajectory, except that this trajectory must exit, at $t=-\infty$, fixed point A and ultimately arrive at extinction hyperplane $q=0$. We know much more, however, in the case of a \textit{very} strong Allee effect, when the basin of attraction of the state $q=q_2$ in the deterministic theory is small.  Here the
noise only needs to create the critical nucleus, see Section II C1. In the language of Eqs.~(\ref{p100}) and (\ref{p110}), the activation trajectory must approach, at $t=+\infty$, fixed point D that involves $q=q_c(x)$ (the critical nucleus) and $p(x)=0$. One can argue that, from there on the population flows to fixed point C (where $q=p=0$) along a relaxation trajectory. The relaxation trajectory does not cost any action (unless the system size $L$ is exponentially large in $K$, see section \ref{caseB} B). In this case the MTE  can be identified, up to a pre-exponent, with the mean time of creation of the critical nucleus. Note that, in this limit, the activation trajectory is again a heteroclinic connection (AD). One can expect that, for a moderately strong Allee effect, the activation trajectory will still
involve a critical nucleus and, therefore, represent a heteroclinic connection AD.

Once the activation trajectory is found, we can obtain the MTE in the leading order of the WKB theory by calculating $S(\textbf{0})$, entering  Eq.~(\ref{rev}), along this trajectory. In scenario I
$S(\mathbf{0})$ is the action along the heteroclinic connection AB.
In the strong-Allee-effect limit of scenario II one has $ \ln (\mu_0 T_e) \simeq K S_0$, where $S_0$ is the action along the heteroclinic connection AD. In both cases we can write $ \ln (\mu_0 T_e) \simeq K {\cal S}$, where
\begin{equation}\label{action}
 {\cal S} = \frac{1}{h} \int_0^L dx \int_{-\infty}^{\infty} dt \,p(x,t) \, \partial_t q(x,t)\,.
\end{equation}
If there are more than one heteroclinic connections between the same pair of fixed points, and obeying the same boundary conditions in space, one should choose the connection which yields the minimum action. Similarly to non-spatial but multi-population systems \cite{EK1,Dykman2,KM,KDM}, the minimum-action trajectory is the most probable path of the population on the way to extinction.
Sections \ref{caseA} and \ref{caseB} present three particular examples of determining the activation trajectories and evaluating the MTE.

As we already mentioned, Hamilton's equations (\ref{p100}) and (\ref{p110}) coincide, upon canonical
transformation $\mathcal{Q}=q e^{-p},\;\mathcal{P}=e^p$, with those derived by Elgart and Kamenev in the framework of a time-dependent WKB approximation \cite{EK1}. There are some differences, however, between our and their formulations of the problem. These difference involve boundary conditions: both in time, and in space. The differences in the boundary conditions in time appear already in the most basic, spatially-independent setting, so let us consider this setting first.

The time-dependent WKB formulation of Ref. \cite{EK1} prescribes, at $t=0$, the initial population size, say $q=q_s$, with an \textit{a priori} unknown momentum. It also prescribes, at a (sufficiently large) final time $t=T$, momentum
${\cal P}=0$ (or, in our variables, $p=-\infty$), with an \textit{a priori} unknown population size.  One needs to find the initial $p$ and the final $q$ from the condition that the action along the resulting trajectory is minimum. $T$ is ultimately sent to infinity \cite{EK1}.

Our WKB formulation differs, first of all, in its prescription of
the final state of the system. In view of Eq.~(\ref{rev}), we demand $q=0$ there.
Furthermore, we know that the activation trajectory must be a heteroclinic connection AB (in scenario I) or $AD$ (in scenario II). This involves a full  knowledge of both $q$ and $p$ at the initial ($t=-\infty$) and final ($t=\infty$) points. Importantly,  the final value of the momentum in this formulation is different from  $p=-\infty$, or ${\cal P}=0$ demanded in Ref. \cite{EK1}.

\begin{figure}[ht]
\includegraphics[width=2.0 in,clip=]{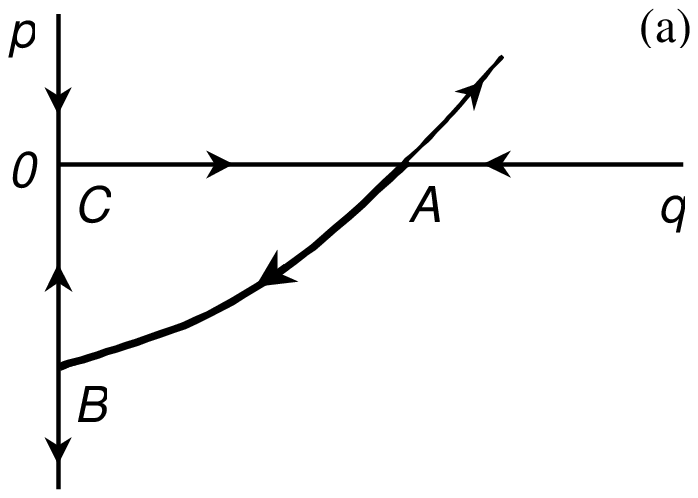}
\includegraphics[width=2.0 in,clip=]{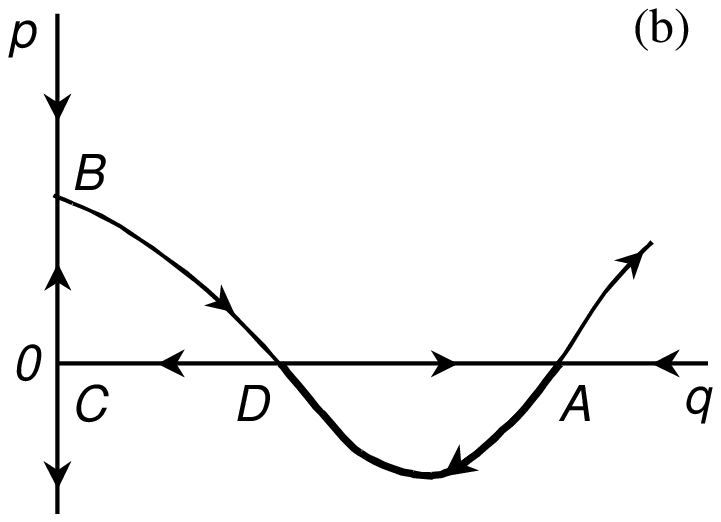}
\caption{Zero-energy trajectories in spatially-independent settings for extinction scenario I (a) and II (b) \cite{EK2,AM2010}. Shown are fixed points A,B and C (a) and A,B,C and D (b). The arrows show stable and unstable manifolds of the corresponding fixed points. The activation trajectories AB (a) and AD (b) are accentuated by thicker lines.} \label{newfig}
\end{figure}

In spite of these differences, the two formulations yield, \textit{in the spatially-independent case}, the same result for the MTE. This happens  because of two reasons. First, the activation trajectory that emerges, at $T\to \infty$, in the time-dependent formulation \cite{EK1} has zero energy, as in our quasi-stationary formulation.  Second, the activation trajectory in the time-dependent formulation is, in general, \textit{not} a heteroclinic connection. Rather, it consists of two (in scenario I) or even three (in scenario II) separate heteroclinic connections  \cite{AM2010,EK2}, see Fig. \ref{newfig}. One of them coincides with trajectory AB or AD (for scenarios I or II, respectively) that the quasi-stationary theory predicts. The other segments go along either $q=0$ or $p=0$ lines and therefore do not contribute to the action. [One can even argue that the last segment ultimately reaches $p=-\infty$, see Fig. \ref{newfig}.]  An advantage of the quasi-stationary theory, especially in numerical calculations,
is that the non-contributing segments of the trajectory are excluded from the start.

For spatially-dependent systems the differences between the two formulations may become irreconcilable. Consider, for example, scenario I in the case of absorbing boundaries.  Here the $x$-independent momentum $p(x)=-\infty$, postulated as the final state in Ref. \cite{EK1}, does not obey the zero boundary conditions $p(0,t)=p(L,t)=0$, and so it cannot possibly be a correct final state.

\section{Population extinction: Scenario I}
\label{caseA}

\subsection{Universal limit}
\label{universal}

Here we consider, as an example, the spatio-temporal SIS model. To render our results a broader relevance, we assume from the start that the basic reproduction number $R_0$ is only slightly larger than $1$: $R_0=1+\delta$, where $0<\delta \ll 1$. In this limit both $q$ and $p$ scale as $\delta$, and
on-site Hamiltonian (\ref{onsiteH}) reduces to
\begin{equation}\label{univ}
    H_0(q,p)\simeq \mu_0 q p (p-q +\delta)\,.
\end{equation}
This on-site Hamiltonian describes, in WKB approximation, a broad class of population models (that do not exhibit Allee effect) close to their transcritical bifurcation at $\delta=0$ \cite{AKM2,AM2010,EK2}.
Notice that, at $\delta \ll 1$, the on-site dynamics exhibits critical slow-down: the characteristic on-site relaxation time becomes $1/(\mu_0\delta)$. As a result, the validity of the continuous description in space here demands $D_0\gg \mu_0 \delta$: a much softer
criterion than $D_0\gg \mu_0$.

Let us define the characteristic diffusion length $l=[D/(\mu_0 \delta)]^{1/2}$ and introduce rescaled population size $Q=q/\delta$,  momentum $P=p/\delta$,  spatial coordinate $\tilde{x}=x/l$, and time $\tilde{t}=\mu_0 \delta\,t$. Upon this rescaling one observes that the second term
in the square brackets in Eq.~(\ref{H21}) is of next order in $\delta$ compared to the rest of terms, and should be neglected. The resulting Hamiltonian density is parameter-free,
\begin{equation}\label{triangle}
    w=QP (P-Q+1)+P \partial_x^2 \,Q\,.
\end{equation}
Here and in the following we drop the tildes everywhere except in the rescaled system size $\tilde{L}=L/l$. Action (\ref{action}) becomes
\begin{equation}\label{actionA}
 {\cal S} (\mathbf{0}) = \frac{\delta^2 l}{h}\,s_A(\tilde{L})\,,
\end{equation}
where
\begin{equation}\label{reducedaction}
s_A(\tilde{L}) = \int_0^{\tilde{L}} dx \int_{-\infty}^{\infty} dt \,P(x,t) \, \partial_t Q(x,t)
\end{equation}
is the rescaled action. The rescaled Hamilton's equations are
\begin{eqnarray}
  \partial_t Q  &=& 2Q P+Q-Q^2+\partial_x^2 Q\,, \label{eqp10}\\
  \partial_t P &=& 2QP-P-P^2-\partial_x^2 P\,. \label{eqp20}
\end{eqnarray}
The same WKB equations can be obtained if one approximates, at small $\delta$, the original master equation by the (functional) Fokker-Planck equation, see Appendix C. This is not surprising, as the validity of the Fokker-Planck approximation demands, in addition to $n_i\gg 1$, that the probability distribution $P(\mathbf{n},t)$  be a slowly varying function of $\mathbf{n}$. The latter condition boils down to condition $|p(x,t)|\ll 1$ that has been used in deriving Eq.~(\ref{univ}). We emphasize that, far from the bifurcation point, the Fokker-Planck approximation in general breaks down, whereas the WKB approximation still holds, in most of the phase space. An example is considered in section \ref{nonuniversal}.

To calculate the rescaled action $s_A$, which only depends on $\tilde{L}$, we need to find a heteroclinic
connection AB. Deterministic steady state $Q=Q_s(x)$, $P=0$, corresponding to fixed point A, is given by the non-trivial solution of equation
\begin{equation}\label{instate}
    Q^{\prime\prime}(x)+Q-Q^2=0\,,
\end{equation}
whereas extinction state $Q=0$, $P=P_{e}(x)$ corresponds to the non-trivial solution of equation
\begin{equation}\label{finstate}
    P^{\prime\prime}(x)+P+P^2=0\,.
\end{equation}
For periodic boundary conditions,  see Fig.~\ref{A2}a,  we obtain $x$-independent solutions: $Q_s(x)=1$, $P_{e}(x)=-1$. As a result, the ``extinction instanton" is $x$-independent for any system size $L$, and one can use the well known one-site WKB results \cite{AKM2,Dykman2,KM,EK2}. The instanton is described, at any $x$, by the equation $P-Q+1=0$. Rescaled action (\ref{reducedaction}) along the extinction instanton is equal to $s_A=\tilde{L}/2$. Then, using Eq.~(\ref{actionA}), we find
\begin{equation}\label{MTE1}
    \ln (\mu_0 \delta T_e)\simeq K S(\mathbf{0})=\frac{K \delta^2 L}{2 h}=\frac{N K\delta^2}{2}
\end{equation}
which is the one-site result times $N$, as expected. The one-site result
for the MTE, $T_e^{(0)}$, is actually known with a higher accuracy -- including a pre-exponential factor \cite{AM2010}:
$$
\mu_0 \delta \,T_e^{(0)} \simeq \sqrt{\frac{2 \pi}{K \delta^2}}\, \exp\left(\frac{K \delta^2}{2}\right)\,.
$$
Therefore, for $L\ll l$, we obtain a more accurate result for $T_e$:
\begin{equation}\label{MTE10}
    \mu_0 \delta \, T_e \simeq \sqrt{\frac{2 \pi}{N K \delta^2}}\, \exp\left(\frac{N K \delta^2}{2}\right)\,.
\end{equation}
as all of the system can be considered here as a single site. Equation~(\ref{MTE10})  holds when
$L\ll l$ and $N K \delta^2 \gg 1$.

Now let us consider a more interesting case of absorbing boundaries: $Q(0,t)=Q(\tilde{L},t)=P(0,t)=P(\tilde{L},t)=0$ (see Appendix A). We notice that, if Eq.~(\ref{instate}) has a nontrivial solution $Q_0(x)$, then Eq.~(\ref{finstate}) has a non-trivial solution $-Q_0(x)$.
Now, Eq.~(\ref{instate}) has a unique nontrivial solution $Q_s(x)$, corresponding to an established population, if $\tilde{L}>\tilde{L}_c=\pi$ or, in dimensional units, $L>L_c=\pi [D/(\mu_0\delta)]^{1/2}$, see~Eq.~(\ref{Lc}). (Note that $L_c=\pi l$ here.) In this case Eq.~(\ref{finstate}) has a nontrivial solution $P_e(x)=-Q_s(x)$ (fixed point B).

\begin{figure}[ht]
\includegraphics[width=2.3 in,clip=]{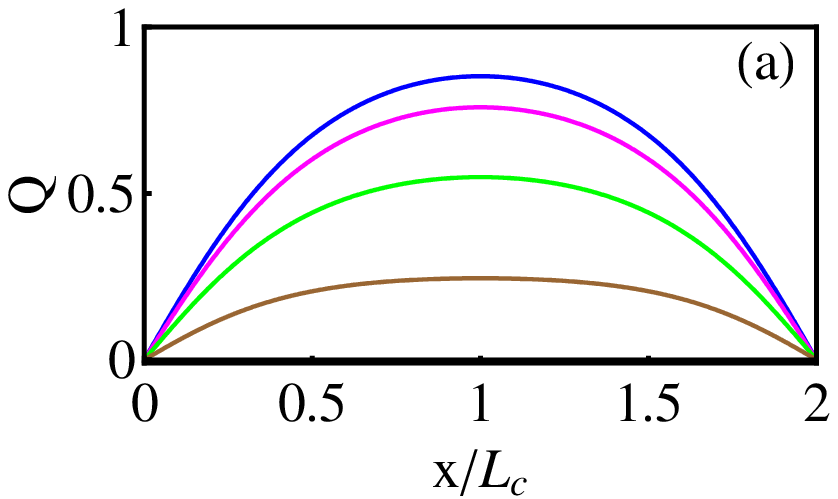}
\includegraphics[width=2.3 in,clip=]{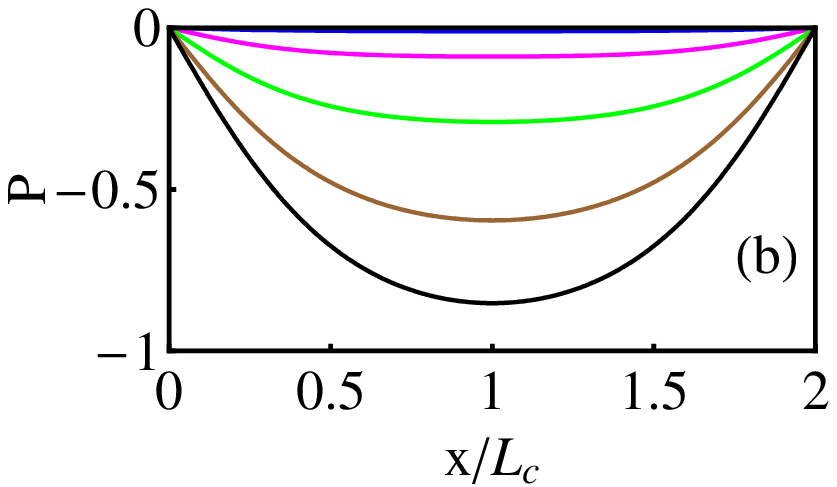}
\caption{(color online) Numerically computed extinction instanton for scenario I (no Allee effect) close to bifurcation point $\delta=0$. The rescaled system length is $L/L_c=2$.
Shown, after 500 iterations of the Elgart-Kamenev numerical algorithm (see text), are spatial profiles of rescaled population size $Q$ (a) and rescaled momentum $P$ (b) at numerical times $0$, $3$, $5$, $7$ and $20$ (from top to bottom). The time interval used for iterations was $0<t<T$ with $T=50$.} \label{numinstanton}
\end{figure}

\begin{figure}[ht]
\includegraphics[width=2.4 in,clip=]{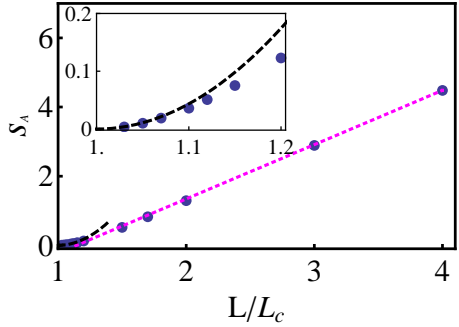}
\caption{(color online)  Rescaled action, Eq.~(\ref{reducedaction}), determining $\ln(\mu_0T_e)$, versus the rescaled system size $\tilde{L}=L/L_c$ for scenario I (no Allee effect) close to the bifurcation point $\delta=0$. Symbols: results obtained with the Elgart-Kamenev numerical algorithm (see text). Dashed line: asymptote $s_A \simeq (9\pi^3/64) (\tilde{L}-1)^2$ for $0<\tilde{L}-1\ll 1$. Dotted line: asymptote  $s_A \simeq (\pi/2) \tilde{L}-1.8$ for $\tilde{L}\gg 1$. Inset: a blowup close to $\tilde{L}=1$.} \label{numericalA}
\end{figure}

Now we need to find a heteroclinic connection AB. To our knowledge, this cannot be done analytically for arbitrary $L>L_c$: even for relatively simple universal Hamiltonian (\ref{triangle}). To solve the problem
numerically, we modified, and implemented in ``Mathematica", the algorithm suggested by Elgart and Kamenev \cite{EK1}. The algorithm iterates Eq.~(\ref{eqp10}) forward in time and Eq.~(\ref{eqp20}) backward in time. It does not involve shooting and
avoids, because of the backward integration in time, the short-wavelength numerical instability caused by the presence of negative diffusion in Eq.~(\ref{eqp20}). In every iteration of $Q(x,t)$ one starts, at $t=0$, from $Q=Q_s(x)$ and solve Eq.~(\ref{eqp10}),  with zero boundary conditions  at $x=0$ and $\tilde{L}$, forward in time until a sufficiently long time $T$ is reached. In this calculation the previous iteration for $P(x,t)$ is used.  Then
Eq.~(\ref{eqp20}) for $P$ is solved backward in time starting, at $t=T$, from $P=P_e(x)$ and continuing until  $t=0$. Here the previous iteration for $Q(x,t)$ is used, and zero boundary conditions at $x=0$ and $\tilde{L}$ are enforced. The very first iteration for $P$ is the desired final steady state $P_e(x)$, satisfying the zero boundary conditions at $x=0$ and $\tilde{L}$ and corresponding to fixed point B.  An example of numerically found instanton is shown in Fig.~\ref{numinstanton}. The filled circles in Fig.~\ref{numericalA} show the numerically computed rescaled action $s_A$, see Eq.~(\ref{reducedaction}), versus rescaled system size $L/L_c$.

Approximate analytic solutions are possible in two limits: $L\gg L_c$ and $0<L-L_c\ll L_c$, and we will present now these solutions. (To remind the reader,
there is no established population at $L<L_c.$)

For $L\gg L_c$, $Q_s(x)$ and $P_e(x)=-Q_s(x)$ are close to
$1$ and $-1$, respectively, everywhere except in boundary layers of width ${\cal O}(1)$ at $x=0$ and $\tilde{L}$. Correspondingly, the extinction instanton is very close (up to corrections exponentially small in $\tilde{L}$) to the one-site instanton $P-Q+1=0$ everywhere except in the boundary layers.  As a result, the rescaled action, $s_A = \tilde{L}/2 -{\cal O} (1)$, differs by a term of order unity from the corresponding result for periodic boundary conditions. The ${\cal O}(1)$ correction, that we found numerically, is about $1.8$, and its contribution to $\ln (\mu_0 \delta\, T_e)$ is relatively large.   The asymptote $s_A = (\pi/2) (L/L_c)-1.8$ is shown in Fig.~\ref{numericalA}. Surprisingly, it works well already at quite small values of $L/L_c-1$. The next-order correction would come from the gradient corrections to the zero boundary conditions in space for $q$ and $p$, see Eqs.~(\ref{e2}), (\ref{e3}) and (\ref{e4}). The expected correction to $s_A$ is ${\cal O}(\mu_0 \delta/D_0)^{1/2}\ll 1$. However, by virtue of Eq.~(\ref{actionA}),  this correction contributes factor ${\cal O}(K \delta^2)$ to  $\ln (\mu_0 \delta \,T_e)$. This contribution is of order of the one-site result for $\ln (\mu_0 \delta \, T_e)$ and may still be large. That is, for $L\gg L_c$  the leading contribution to  $\ln (\mu_0 \delta \,T_e)$ scales as $L$, the subleading contribution scales as $L_c\ll L$, and the sub-subleading
contribution scales as $h\ll L_c$, the latter one ``remembers" the lattice formulation of the problem. The sub-subleading correction can be calculated numerically using the modified boundary conditions. Note that criterion $L\gg L_c$ becomes stringent as the bifurcation point $\delta=0$ is approached, and $L_c$ diverges.

Now consider the limit of $0<L-L_c\ll L_c$. We start with a perturbative calculation of $Q_s(x)$ and $P_e(x)$. Let $\varepsilon=\tilde{L}-\tilde{L}_c=\tilde{L}-\pi \ll 1$. For $Q_s(x)$ we can write
$$
Q_s(x)\simeq a_0+a_1 \sin x + b_1 \cos x + a_2 \cos 2x\,,
$$
where $a_1 \sim \varepsilon$, whereas $a_0\sim b_1 \sim a_2 \sim \varepsilon^2$. Plugging this ansatz into Eq.~(\ref{instate}), we obtain $a_0=a_1^2/2$ and $a_2 =a_1^2/6$.
Boundary condition $Q(0)=0$ yields $b_1=-2 a_1^2/3$. Now we demand $Q(\tilde{L})\equiv Q(\pi+\varepsilon)=0$. Expanding
this condition at small $\varepsilon$, we obtain
\begin{equation}\label{a(L)}
    a_1=\frac{3 \varepsilon}{4}=\frac{3}{4}\left(\tilde{L}-\tilde{L}_c\right)= \frac{3}{4}\left(\tilde{L}-\pi\right)\,.
\end{equation}
The bifurcation of the steady-state solutions, both $Q_s(x)$ and $P_e(x)$, at $\tilde{L}=\tilde{L}_c=\pi$ is, therefore, transcritical. The final result for $Q_s(x)=-P_e(x)$, up to $\varepsilon^2$, is
\begin{equation}\label{qinit}
    Q_s(x)\simeq \frac{9 \varepsilon^2}{32}+\frac{3 \varepsilon}{4} \sin x - \frac{3 \varepsilon^2}{8} \cos x +\frac{3\varepsilon^2}{32} \cos 2 x\,.
\end{equation}
Now let us solve perturbatively Hamilton's equations~(\ref{eqp10}) and (\ref{eqp20}).
Shrinking the coordinate $x$,
$$
\frac{\pi}{\pi+\varepsilon}\, x \to x\, ,
$$
we rewrite these equations as
\begin{eqnarray}
  \partial_t Q  &=& 2QP+Q-Q^2+\left(1-\frac{2\varepsilon}{\pi}+\dots\right)\,\partial_x^2Q\,,\label{eqp30}\\
  \partial_t P &=&2QP-P-P^2-\left(1-\frac{2\varepsilon}{\pi}+\dots\right)\, \partial_x^2 P\,, \label{eqp40}
\end{eqnarray}
where dots denote higher order terms in $\varepsilon$. The problem is now defined on the interval $0\leq x\leq \pi$. We seek perturbative solutions in the form
\begin{eqnarray}
  Q(x,t) &=& \varepsilon u(x, \varepsilon t)+\varepsilon^2 u_1(x, \varepsilon t)+ \dots\,, \nonumber\\
  P(x,t) &=& \varepsilon v(x, \varepsilon t)+\varepsilon^2 v_1(x, \varepsilon t)+ \dots\,.\nonumber
\end{eqnarray}
In the first order in $\varepsilon$ we obtain equations
\begin{equation}\label{firstorder}
    \partial_x^2 u+u=0 \quad \mbox{and} \quad \partial_x^2 v+v=0\,.
\end{equation}
Their solutions, obeying zero boundary conditions at $x=0$ and $\pi$, are
\begin{equation}
u(x,\tau)=a(\tau)\,\sin x\, ,\quad v(x,\tau)=b (\tau)\,\sin x\,,
\label{eqp50}
\end{equation}
where $a(\tau)$ and $b(\tau)$ are yet unknown functions of the slow time $\tau=\varepsilon t$.
In the second order in $\varepsilon$ we obtain
\begin{eqnarray}
\!\!\!\!\!\!\!\!\!\!\!\partial_x^2 u_1+u_1&=&\left(\frac{da}{d\tau}-\frac{2a}{\pi}\right) \sin x +\left(a^2-2ab\right) \sin^2 x,
\label{eqp90}\\
\!\!\!\!\!\!\!\!\!\!\!\partial_x^2 v_1+v_1&=&-\left(\frac{db}{d\tau}+\frac{2b}{\pi}\right) \sin x -\left(b^2-2ab\right) \sin^2 x,
\label{eqp100}
\end{eqnarray}
subject to zero boundary conditions at $x=0$ and $\pi$. The solvability conditions for Eqs.~(\ref{eqp90}) and~(\ref{eqp100}) yield the following equations for $da/d\tau$ and $db/d\tau$:
\begin{eqnarray}
\frac{da}{d\tau}&=&\frac{2a}{\pi}-\frac{8}{3\pi} \left(a^2-2ab\right)\,,
\label{eqp140}\\
\frac{db}{d\tau}&=&-\frac{2b}{\pi}-\frac{8}{3\pi} \left(b^2-2ab\right)\,,
\label{eqp150}
\end{eqnarray}
These are Hamilton's equations for generalized coordinate $a$ and momentum $b$. Hamiltonian
\begin{equation}
{\cal H}(a,b)=\frac{8}{3\pi} a  b \left(b-a+\frac{3}{4}\right)
\label{eqp180}
\end{equation}
is of the same type as universal \textit{on-site} Hamiltonian (\ref{univ}). The extinction instanton obeys
$b=a-3/4$, and we find
\begin{eqnarray}
a&=&\frac{3}{4\, \left(1+e^{2\varepsilon t/\pi}\right)}\, ,
\label{eqp200}\\
b&=&-\frac{3}{4\, \left(1+e^{-2\varepsilon t/\pi}\right)}\, .
\label{eqp210}
\end{eqnarray}
One can also easily find the second-order corrections $u_1$ and $v_1$ from Eqs.~(\ref{eqp90}) and~(\ref{eqp100}), but we will not present these formulas here.
Now we calculate, in the leading order in $\varepsilon$, the rescaled action~(\ref{reducedaction}) along the extinction instanton:
\begin{eqnarray}
  s_A &=& \int_0^{\pi+\varepsilon} dx \int_{-\infty}^{\infty} dt \,p(x,t) \, \partial_t q(x,t) \nonumber \\
  &\simeq&
  \varepsilon^2 \int\limits_0^{\pi} dx  \, \sin^2 x \int\limits_{-\infty}^{+\infty}d\tau \,b\,\frac{da}{d\tau} \nonumber\\
  &=& \frac{\pi\varepsilon^2}{2} \int\limits_{3/4}^{0}da
  \left(a-\frac{3}{4}\right) = \frac{9\pi\varepsilon^2}{64}\,. \label{s10}
\end{eqnarray}
This asymptote is shown in Fig.~\ref{numericalA}.
Finally, we use Eqs.~(\ref{actionA}) and (\ref{s10}) to find the logarithm of the MTE:
\begin{equation}\label{MTE2}
    \ln (\mu_0 \delta\,T_e)\simeq K S(\mathbf{0})=\frac{9 \pi^2 K \delta^2 L_c}{64 h} \left(\frac{L}{L_c}-1\right)^2\,.
\end{equation}
This result is valid, for $L-L_c \ll L_c$, when it is much greater than unity. This holds for sufficiently large $K$ or fast migration.

\subsection{Extinction of particles undergoing reactions $A \to 2A$ and $2A \to 0$}
\label{nonuniversal}

Here we use WKB approximation to revisit the problem of extinction of particles $A$ which participate in two on-site reactions: branching $A \to 2A$ and annihilation $2A \to 0$, with rate coefficients $\mu_0$ and $\mu_0/K$, respectively, and $K \gg 1$.  Although still exhibiting scenario I of extinction, the on-site Hamiltonian for this model is irreducible and does not belong to the universality class considered in the previous subsection.  In the spatially-independent formulation, the Fokker-Planck approximation does not apply for the evaluation of the MTE \cite{Kessler}. All this is because of the absence of linear decay process $A\to 0$ (or of the linear in $n$ small-$n$ asymptote of the death rate). Here the extinction instanton, in the spatially-independent  setting, does approach, at $t\to \infty$, infinite momentum $p=-\infty$ (which, in our variables, is the ``extinction momentum").

In spite of its degeneracy, this model is quite popular. Its spatially-independent version was investigated in many papers, see \textit{e.g.} Refs. \cite{Oppenheim,EK1,Kessler,AM2}, whereas  the spatial version was considered in Ref. \cite{EK1}. Our objectives here are three-fold. First, we use this example to illustrate the advantages of canonical variables $\mathcal{Q}$ and $\mathcal{P}$ (that arise naturally in the probability generating function formalism \cite{EK1,KMS}). Second, we show that, when the system size $L$ is
only slightly above $L_c$, this ``irreducible" model does reduce to the universality class considered in Sec. \ref{caseA}. Third, we use this example to compare our results with those of Elgart and Kamenev \cite{EK1}.

We start from Eq.~(\ref{H20}) for the Hamiltonian functional. The density $w$ has the form of Eq.~(\ref{H21}), whereas  the on-site Hamiltonian for processes $A \to 2A$ and $2A \to 0$ is the following:
\begin{equation}\label{H01B}
    H_0(q,p)= \mu_0 q \left(e^p-1\right)+\frac{1}{2} \,\mu_0 q^2 \left(e^{-2p}-1\right)\,.
\end{equation}
Define characteristic diffusion length $l=(D/\mu_0)^{1/2}$.
Introducing rescaled coordinate $\tilde{x}=x/l$ and time $\tilde{t}=\mu_0 t$, we arrive at a parameter-free Hamiltonian
with density
\begin{eqnarray}
  w &=& q \left(e^p-1\right)+\frac{1}{2} \,q^2 \left(e^{-2p}-1\right) \nonumber\\
  &-& \partial_x q\, \partial_x p+q \left(\partial_x p\right)^2\,,
 \label{H02B}
\end{eqnarray}
where we have dropped the tildes. It is advantageous to make a canonical transformation from $q$ and $p$ to  $\mathcal{Q}=q e^{-p}$ and $\mathcal{P}=e^p-1$ (the shift by $1$ in $\mathcal{P}$ preserves the deterministic line at $\mathcal{P}$=0). The new Hamiltonian density becomes
\begin{equation}\label{densityapB}
\mathcal{W}=\mathcal{Q} \mathcal{P}\left[\mathcal{P}-\mathcal{Q}+1 -(1/2) \mathcal{Q} \mathcal{P} \right] - \partial_x \mathcal{Q}\, \partial_x \mathcal{P}\,,
\end{equation}
and the Hamilton's equations are \cite{EK1}
\begin{eqnarray}
  \partial_t \mathcal{Q}  &=& 2\mathcal{Q} \mathcal{P}+\mathcal{Q}-\mathcal{Q}^2 - \mathcal{Q}^2 \mathcal{P}+\partial_x^2 \mathcal{Q}\,, \label{10B}\\
  \partial_t \mathcal{P} &=& \mathcal{Q}\mathcal{P}^2+2 \mathcal{Q}\mathcal{P}-\mathcal{P}-\mathcal{P}^2-\partial_x^2 \mathcal{P}\,. \label{20B}
\end{eqnarray}
Extinction action (\ref{action}) becomes
\begin{equation}\label{actB}
 {\cal S} (\mathbf{0}) = \frac{l}{h}\,s(\tilde{L})\,,
\end{equation}
where $\tilde{L}=L/l$ is the rescaled system size, and
\begin{equation}\label{redactB}
s(\tilde{L}) = \int_0^{\tilde{L}} dx \int_{-\infty}^{\infty} dt \,\mathcal{P}(x,t) \, \partial_t \mathcal{Q}(x,t)
\end{equation}
is the rescaled action. To calculate $s(\tilde{L})$ we need to find an instanton-like activation trajectory that exits from fixed point A  at $t=-\infty$ and asymptotically approaches the proper extinction state at $t=\infty$.  Fixed point A, corresponding to $\mathcal{Q}=\mathcal{Q}_s(x)$, $\mathcal{P}=0$, is given by the non-trivial solution of equation
\begin{equation}\label{instateB}
    \mathcal{Q}^{\prime\prime}(x)+\mathcal{Q}-\mathcal{Q}^2=0\,,
\end{equation}
with boundary conditions $\mathcal{Q}(0)=\mathcal{Q}(\tilde{L})=0$.
In its turn, the proper extinction state $\mathcal{Q}=0$, $\mathcal{P}=\mathcal{P}_{e}(x)$ is given by the non-trivial solution of equation
\begin{equation}\label{finstateB}
    \mathcal{P}^{\prime\prime}(x)+\mathcal{P}+\mathcal{P}^2=0\,,
\end{equation}
with $\mathcal{P}(0)=\mathcal{P}(\tilde{L})=0$. Interestingly,  the equations and boundary conditions for $\mathcal{Q}_s(x)$ and $\mathcal{P}_{e}(x)$ coincide with those for $Q_s(x)$ and  $P_e(x)$ for the universal model of scenario I, see Eqs.~(\ref{instate}) and (\ref{finstate}).
In particular, equality $\mathcal{P}_e(x)=-\mathcal{Q}_s(x)$ holds.

\begin{figure}[ht]
\includegraphics[width=2.3 in,clip=]{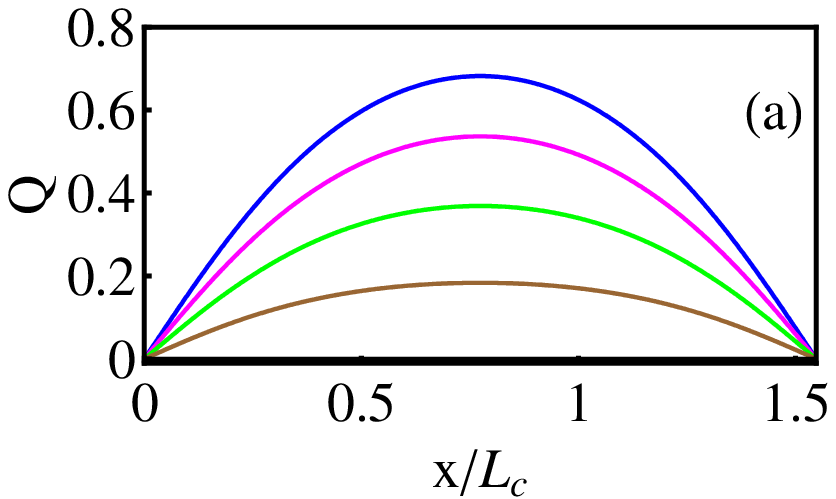}
\includegraphics[width=2.3 in,clip=]{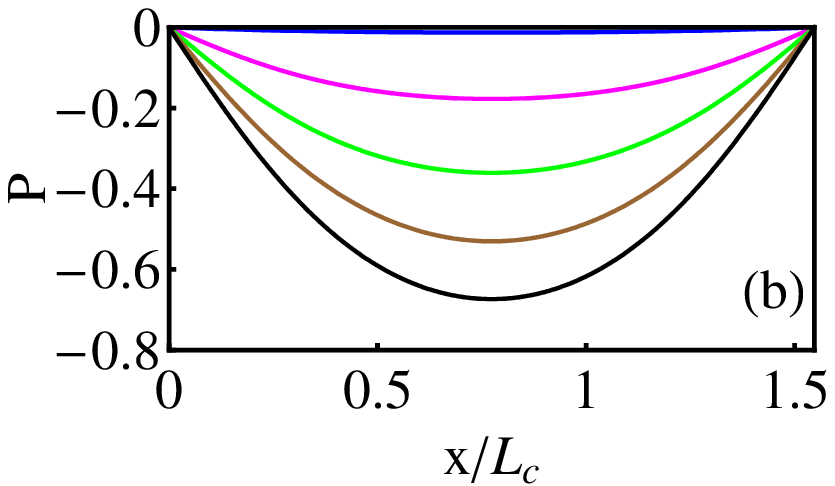}
\caption{(color online) Numerically computed extinction instanton for processes $A\to 2A$ and $2A\to 0$. The rescaled system length is $L/L_c=1.55$.
Shown, after 400 iterations of the Elgart-Kamenev numerical algorithm (see subsection \ref{universal}), are spatial profiles of $\mathcal{Q}$ (a) and $\mathcal{P}$ (b) at numerical times $0$, $5$, $7$, $9$ and $50$ (from top to bottom).  The time interval used for the iterations was $0<t<T$ with $T=60$.} \label{nodecay}
\end{figure}

We solved the problem numerically using the Elgart-Kamenev algorithm described above. An example of numerically found instanton is shown in Fig.~\ref{nodecay}. The time-dependent solution of this problem is different from that of the universal model, except when
the rescaled system size $\tilde{L}$ only slightly exceeds the rescaled critical size for the established population, $\tilde{L}_c=\pi$. Here $|\mathcal{Q}|\ll 1$ and $|\mathcal{P}| \ll 1$ for all $0<x<\tilde{L}$ and $-\infty<t<\infty$, and we can neglect
the last term in the square brackets of Hamiltonian density (\ref{densityapB}), thus arriving
at universal Hamiltonian (\ref{triangle}) considered in the previous subsection. As a result,
the rescaled action in this case is again $s \simeq (9\pi^3/64) (L/L_c-1)^2$, see Eq.~(\ref{s10}).
Surprisingly, this result agrees with that obtained by Elgart and Kamenev close to $L=L_c$, see  Eq.~(45) of Ref. \cite{EK1}. The reason for this agreement is unclear, as Elgart and Kamenev do not mention any boundary conditions for $\mathcal{P}$ at  $x=0$ and $x=L$.

\begin{figure}[ht]
\includegraphics[width=2.3 in,clip=]{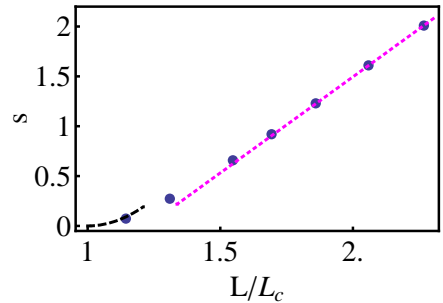}
\caption{(color online) Rescaled action, Eq.~(\ref{redactB}), versus the rescaled system size $\tilde{L}=L/L_c$ for the processes $A\to 2A$ and $2A\to 0$. Symbols: results obtained with the Elgart-Kamenev numerical algorithm, see section \ref{universal}. Dashed line: asymptote $s \simeq (9\pi^3/64) (\tilde{L}-1)^2$ for $0<\tilde{L}-1\ll 1$. Dotted line: asymptote  $s \simeq 2 \pi (1-\ln 2) \tilde{L}-2.36$ for $\tilde{L}\gg 1$.} \label{nolineardecay}
\end{figure}

For $L\gg L_c$ functions $\mathcal{Q}_s(x)$ and $\mathcal{P}_e(x) = -\mathcal{Q}_s(x)$ are close
to $1$ and $-1$, respectively, except in the boundary layers at $x=0$ and $\tilde{L}$. As a result, the extinction instanton is close to the one-site instanton $\mathcal{Q}=2 (\mathcal{P}-1)/(\mathcal{P}+2)$ everywhere except in the boundary layers, and we obtain $s \simeq 2 \pi (1-\ln 2) (L/L_c)-2.36$, see Fig.~\ref{nolineardecay}. Factor $2 (1-\ln 2)$ comes from the solution of the one-site problem \cite{Oppenheim,EK1,Kessler,AM2}. The leading term, proportional to $L$, coincides with that obtained by Elgart and Kamenev~\cite{EK1}. The numerically found offset $2.36$  also agrees, up to $1$ percent, with their result. We can only explain this agreement (and the agreement in our analytic results at $L/L_c-1 \ll 1$, reported above) by assuming that Elgart and Kamenev did impose correct spatial boundary conditions at the edges of the system, $x=0$ and $x=L$. But then the final state $\mathcal{P}(x)$ in their
calculations must have been $\mathcal{P}=\mathcal{P}_e(x)=-\mathcal{Q}_s(x)$, and not  $\mathcal{P}=0$
as they claim.

\section{Population extinction: Scenario II, very strong Allee effect}
\label{caseB}

Here we consider the three reactions $A\to 0$ and  $2A \rightleftarrows 3A$. They are described, in WKB approximation, by Hamilton's equations ~(\ref{p100}) and (\ref{p110}) with on-site
Hamiltonian
\begin{equation}\label{hamilex2}
H_0(q,p)=\mu_0\left(\frac{q^3}{\gamma}+q\right)(e^{-p}-1)+\frac{2\mu_0 q^2}{\gamma}(e^p-1)\,.
\end{equation}
In the following we only deal with a very strong Allee effect in a system with periodic boundary conditions, see Figs.~\ref{Bstrong1} and~\ref{Bstrong2}. Here the noise-driven population extinction requires a large fluctuation that creates, at $L>L_c$, critical nucleus $q=q_c(x)$.
The ``nucleation instanton" (that is, a heteroclinic connection AD) can be found by solving Eqs.~(\ref{p100}) and (\ref{p110}) with periodic boundary conditions in space for $q(x,t)$ and $p(x,t)$, conditions
$q(x,t \to -\infty)=q_2=1+\delta$ and  $p(x,t \to -\infty)=0$, and conditions $q(x,t \to +\infty)=q_c(x)$ and $p(x,t \to +\infty)=0$, where $0<\delta \ll 1$.

\subsection{Small and moderately large systems}
\label{single}

For $\delta\ll 1$ the on-site deterministic dynamics is close to the saddle-node bifurcation. Here the unstable and stable fixed points, $q_1=1-\delta$ and $q_2=1+\delta$ are both close to $1$,
whereas the momentum $p$ on the activation trajectory scales as $\delta^2$. Expanding on-site Hamiltonian (\ref{hamilex2}) at small $p$ and $q-1$ we arrive at  \cite{AM2010}:
\begin{equation}\label{univ1}
    H_0(q,p)\simeq 2\mu_0 p \,\left(p +\frac{\delta^2-\Delta q^2}{2}\right)\,,
\end{equation}
where $\Delta q = q-1$. This on-site WKB Hamiltonian, considered already in Ref. \cite{Dykman3}, describes a host of spatially-independent overdamped physical systems which exhibit activated escape close to a saddle-node bifurcation. Furthermore, the exact destination of the escape process (whether it is population extinction \cite{AM2010}, population explosion \cite{EK1,MS}, or a switch to another metastable state \cite{EscuderoKamenev}) is of no importance: it is decay of  metastable state $q=q_2$ which is the kinetic bottleneck of the process.  Note that, at $\delta\ll 1$, the fast-migration criterion in the spatial problem becomes $D_0\gg \mu_0\delta$, as in scenario I.

Let us define characteristic diffusion length $l=[D/(2\mu_0 \delta)]^{1/2}$ and
introduce rescaled population size $Q=\Delta q/\delta$, momentum $P=p/\delta^2$,  spatial coordinate $\tilde{x}=x/l$ and time $\tilde{t}=2\mu_0 \delta\,t$. At $\delta\ll 1$ the second term
in the square brackets in Eq.~(\ref{H21}) is again negligible. The resulting (parameter-free) Hamiltonian density can be written as
\begin{equation}\label{parabolic}
    w=P \left[P-U^{\prime}(Q)+\partial_x^2 Q\right]\,,
\end{equation}
where $U(Q)=-Q/2+Q^3/6$ is the effective potential. Here and in the following we drop the tildes
everywhere except in the rescaled system size  $\tilde{L}=L/l$. Action (\ref{action}) becomes
\begin{equation}\label{actionB}
 {\cal S} = \frac{\delta^3 l}{h}\,s_B(\tilde{L})\,,
\end{equation}
where
\begin{equation}\label{reducedactionB}
s_B(\tilde{L}) = \int_0^{\tilde{L}} dx \int_{-\infty}^{\infty} dt \,P(x,t) \, \partial_t Q(x,t)
\end{equation}
is the rescaled action. The rescaled Hamilton's equations are
\begin{eqnarray}
  \partial_t Q(x,t) &=&2P-U^{\prime}(Q)+\partial_x^2 Q\,,
   \label{twoeqns1a} \\
  \partial_t P(x,t) &=&P U^{\prime\prime}(Q)-\partial_x^2 P\,. \label{twoeqns1b}
\end{eqnarray}
The same WKB equations can be obtained from the Fokker-Planck equation, valid at small $\delta$, see Appendix C.

Hamiltonian~(\ref{parabolic}) almost coincides with the Hamiltonian  considered by Elgart and Kamenev \cite{EK1}. The only difference is that the effective potential in their case was of the opposite sign, as they
considered population explosion rather than extinction. The procedure of finding the activation trajectory  (heteroclinic connection AD) is identical in the two cases. It is based on the following important property of zero-energy flows in this class of Hamiltonians: if the quantity
$F(x,t)=P-U^{\prime}(Q)+\partial_x^2 Q$ vanishes,
at some time, for all $x$, then it vanishes at all times. This property can be easily proved by calculating $\partial_t F(x,t)$ and using Eqs.~(\ref{twoeqns1a}) and~(\ref{twoeqns1b}). Since in our problem $Q(x,t=-\infty)=1$ and $P(x,t=-\infty)=0$, equality $F(x,t)=0$ does hold. It immediately follows that $P=\partial_t Q$,  and
\begin{equation}\label{instB}
    \partial_t Q=U^{\prime}(Q)-\partial_x^2 Q
\end{equation}
on the activation trajectory. Equation~(\ref{instB}) is a time-reversed version of the deterministic equation
\begin{equation}\label{mf}
    \partial_t Q=-U^{\prime}(Q)+\partial_x^2 Q\,.
\end{equation}
Using the relation $P=\partial_t Q$ in Eq.~(\ref{reducedactionB}), one obtains $s_B=\Delta \tilde{{\cal F}}$, where
$\Delta \tilde{{\cal F}}$ is the difference between the final (at $t=+\infty$) and initial (at $t=-\infty$) values of the rescaled Ginzburg-Landau free energy functional, \textit{cf.} Eq.~(\ref{freeenergy}):
\begin{equation}\label{freenergy}
    \tilde{{\cal F}} [Q(x,t)] = \int_0^{\tilde{L}} \,dx\, \left[U(Q)+(1/2)(\partial_x Q)^2\right]\,.
\end{equation}
Note that local identity $F(x,t)=0$ implies an infinite number of integrals of motion. Although their presence looks as a miracle in the WKB formalism, it is a direct consequence of integrability of the stationary Fokker-Planck equation in this case, see Appendix C.

The final state, at $L>L_c$, is the (rescaled) critical nucleus: an $x$-dependent solution of the steady-state equation
\begin{equation}\label{nucleus}
    Q^{\prime\prime}(x)+(1/2)[1-Q^2(x)]=0
\end{equation}
subject to periodic boundary conditions with spatial period $\tilde{L}$. Elgart and Kamenev \cite{EK1} solved this equation, and calculated $\Delta \tilde{{\cal F}}$, in the limit of $L \gg L_c$. We will present the solution for any $L>L_c$. The solution of Eq.~(\ref{nucleus}), up to an arbitrary shift in $x$, can be written as
\begin{equation}\label{nucgen}
   Q_c(x)=c + (b-c) \,\mbox{sn}^2 \left[\frac{2 \mbox{K}(m)\, x}{\tilde{L}}\right]\,.
\end{equation}
Here $b$ and $c$ are two of the three real roots $a({\cal E})>b({\cal E})>c({\cal E})$ of the polynomial ${\cal E}-\xi/2+\xi^3/6$ (the roots are real for $|{\cal E}|<1/3$), $\mbox{sn}(\dots)$ is the Jacobi elliptic function, and $\mbox{K}(m)$ is the complete elliptic integral of the first kind \cite{Stegun}. Furthermore, $m=m({\cal E})=(b-c)/(a-c)$,
and parameter ${\cal E}$ is determined by relation
$$
\frac{4 \sqrt{3} \,\mbox{K}(m)}{\sqrt{a-c}}=\tilde{L}\,.
$$
The $x$-dependent solution (\ref{nucgen}) exists at $\tilde{L}>\tilde{L}_c=2\pi$ or, in dimensional units,  $L>L_c=\pi[2D/(\mu_0 \delta)]^{1/2}$; this is what Eq.~(\ref{Lc1}) predicts at $\delta\ll 1$. Note that $L_c=2\pi l$ here.  At $L>L_c$ solution~(\ref{nucgen}) exhibits a single full spatial oscillation: its spatial period is equal to the rescaled system size $\tilde{L}$. At $L>kL_c$, where $k=2,3, \dots$, this solution coexists with additional solutions having $2,3,\dots, k$ full oscillations.   The $k>1$ solutions, however, yield greater actions than solution~(\ref{nucgen}), and therefore should be ruled out.

\begin{figure}[ht]
\includegraphics[width=1.8 in,clip=]{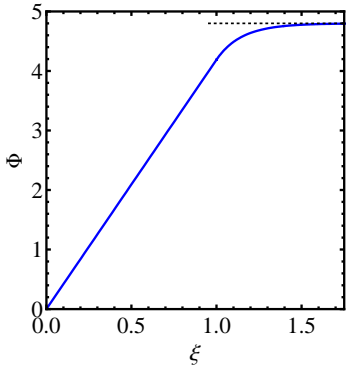}
\caption{(color online) Function $\Phi(\xi)$ determining the dependence of $\ln(\mu_0T_e)$ on the rescaled system size $L/L_c$, see Eq.~(\ref{MTEB}), for populations with a very strong Allee effect. The dotted line is the asymptote $\Phi(\xi\gg 1)=24/5=4.8$.} \label{Phi}
\end{figure}

Evaluating free energy (\ref{freenergy}) for solution (\ref{nucgen}) with the help of ``Mathematica", we obtain $s_B= \Delta \tilde{{\cal F}} =\Phi(L/L_c)$ where function $\Phi(\xi)$ is depicted in Fig. \ref{Phi}.
The logarithm of the MTE is, therefore, approximately equal to
\begin{equation}\label{MTEB}
    \ln(\mu_0 \delta\,T_e)\simeq \frac{K \delta^3 L_c}{2\pi h}\, \Phi\left(\frac{L}{L_c}\right)\,.
\end{equation}
This result is valid when it is much greater than unity. This can be achieved for sufficiently large $K$ and fast migration. Importantly, at $L \gtrsim L_c$ the MTE ceases to grow with system size $L$, so the MTE can be relatively short. Furthermore, for any $L>L_c$ the action spent on creating the critical nucleus is less than the action spent on bringing the population to the $x$-independent unstable state $q=q_1$. Therefore, extinction via the critical nucleus is (exponentially) more probable than via the state $q=q_1$.

What are the asymptotes of this result in three characteristic regions: $L<L_c$, $0<L-L_c\ll L_c$ and $L\gg L_c$? Instead of using asymptotics of the elliptic functions, one can directly solve Eq.~(\ref{nucleus}) in these regions.
At $L<L_c$ the critical nucleus gives way to the $x$-independent unstable solution $Q=-1$. Here we obtain
$\Delta \tilde{{\cal F}}=2\tilde{L}/3$, and
\begin{equation}\label{onesiteB}
    \ln(\mu_0 \delta\,T_e)\simeq\frac{2 L K \delta^3}{3h}=\frac{2 N K \delta^3}{3}\,,
\end{equation}
which is the one-site result \cite{EK1,AM2010} times $N$ as expected. Again, for $L\ll L_c$  we can use
a more accurate one-site result \cite{AM2010} and obtain
\begin{equation}\label{onesiteB10}
    \mu_0 \delta\,T_e \simeq \pi\, \exp\left(\frac{2}{3}\,NK \delta^3\right)\,,
\end{equation}
as the whole system can be considered as a single site. This result is valid when
$L\ll L_c$ and $N K \delta^3 \gg 1$.

At $L=L_c$ a weakly inhomogeneous critical nucleus emerges via a super-critical bifuraction. At $0<L-L_c \ll L_c$ the critical nucleus, in the rescaled variables, is
\begin{equation}\label{weakB}
    Q_{c}(x) \simeq -1+ A \cos \left(\frac{2 \pi x}{\tilde{L}}\right)+\frac{A^2}{4}-\frac{A^2}{12} \cos\left(\frac{4 \pi x}{\tilde{L}}\right)\,,
\end{equation}
where
\begin{equation}\label{AB}
    A\simeq\frac{4 \sqrt{3}}{\sqrt{5}}\,
   \left(\frac{L}{L_c}-1\right)^{1/2}\,.
\end{equation}
Here we obtain
\begin{equation}
  \Delta \tilde{{\cal F}} \simeq \frac{2 \tilde{L}}{3}\left[1-\frac{18}{5}\left(\frac{L}{L_c}-1\right)^2\right] \,,
  \nonumber
\end{equation}
and so
\begin{equation}
     \ln(\mu_0 \delta\,T_e)\simeq \frac{2 N K \delta^3}{3}\,\left[1-\frac{18}{5}\left(\frac{L}{L_c}-1\right)^2\right]\,.
\end{equation}
Finally, at $L\gg L_c$ the critical nucleus can be approximated by its asymptote at $L\to \infty$:
\begin{equation}\label{soliton}
    Q_{c}(x) \simeq 1 - 3 \cosh^{-2}(x/2)\,.
\end{equation}
In this limit, mathematically identical to the one considered by Elgart and Kamenev \cite{EK1} in the context of population explosion, we obtain $s_B=\Delta \tilde{{\cal F}}=24/5$, and so
\begin{equation}\label{MTEnucleus}
    \ln(\mu_0 \delta\,T_e)\simeq \frac{24 K \delta^3 l}{5 h}=\frac{12 K \delta^3 L_c}{5 \pi h}\,.
\end{equation}
Note that criterion $L\gg L_c$ becomes stringent as the bifurcation point $\delta=0$ is approached, and $L_c$ diverges.

Now we see that the asymptotes of $\Phi(\xi)$ are the following:
\begin{equation}\label{deltaF}
\Phi(\xi)=\left\{\begin{array}{lll}
\frac{4\pi}{3} \xi\,,  & \mbox{$\xi<1$}, \\
\frac{4\pi}{3} \xi \left[1-\frac{18}{5}\left(\xi-1\right)^2\right]\,, & \mbox{$0<\xi-1 \ll 1$}, \\
\frac{24}{5}\,, & \mbox{$\xi\gg 1$.}
\end{array}
\right.
\end{equation}

\begin{figure}[ht]
\includegraphics[width=2.3 in,clip=]{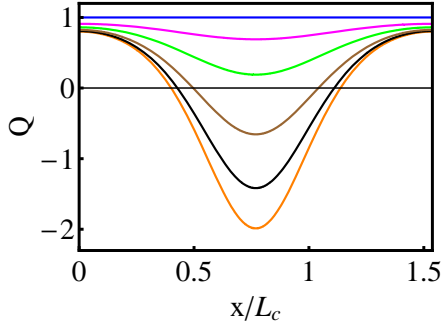}
\caption{(color online) A numerically computed nucleation instanton (heteroclinic connection AD) for scenario II (a strong Allee effect) close to bifurcation point $\delta=0$. The rescaled system length is $L/L_c\simeq 1.54$.
Shown are spatial profiles of the rescaled population size $Q$ at numerical times $0$, $11$, $12$, $13$, $14$ and $20$ (from top to bottom). The last profile is the critical nucleus for this system size.} \label{instanton_B}
\end{figure}

Before concluding this subsection we note that, although we succeeded in calculating action $s_B$ analytically,  the calculation of the nucleation instanton, $Q(x,t)$ and $P(x,t)$, demands (rather simple) numerics.  $Q(x,t)$ is described by the time-reversed version of reaction-diffusion equation (\ref{mf}). Therefore, for a given system size, one can
solve Eq.~(\ref{mf})  with periodic boundary conditions numerically, starting from the critical nucleus (with a slight positive offset) and advancing the solution until it converges close to $Q=1$. The instanton solution is then readily obtained via time reversal, whereas $P(x,t)$ can be found  from
$P(x,t)=\partial_t Q(x,t)$. An example of instanton, found in this way, is depicted in Fig.~\ref{instanton_B}.

\subsection{Very large systems: single versus multiple nucleation}
\label{multiple}

In the previous subsection we evaluated $\ln (\mu_0 \delta\,T_e)$ under condition that the probability of creating more than one critical nucleus during the traverse time of the deterministic extinction fronts through the population is negligible. The rest of parameters being fixed, this condition is always satisfied at sufficiently large $K$. If we instead fix $K\gg 1$ and increase $L$, we will arrive at the regime when additional critical nuclei typically appear while the extinction fronts still run through the population. The present subsection deals with this regime. Importantly, the assumption of quasi-stationarity, see Eqs.~(\ref{qsd1}) and (\ref{P0}), does not hold in this regime, except for the purpose of calculation of the rate of formation of a single critical nucleus in the phase $q_2=1+\delta$. The latter is given, at $L\gg L_c$ and $t\gg 1/(\mu_0\delta)$, by Eq.~(\ref{MTEnucleus}) that we rewrite here, up to pre-exponential factors, as
\begin{equation}\label{rate10}
    \frac{1}{T_e} \propto \mu_0 \delta \,  e^{-KS_0}\,,\;\;\;\mbox{where}\;\;\;S_0= \frac{12\delta^3 L_c}{5 \pi h}\,.
\end{equation}
As $1/T_e$ is exponentially small, the nucleation acts are rare, and we can assume that they are statistically independent and Poisson-distributed.  This implies that, in sufficiently large systems, $L\gg L_c$,  nucleation rate (\ref{rate10}) includes a pre-exponential factor proportional to $L$ (that we did not care about previously but must account for now). Correspondingly, the nucleation rate \textit{density} $\rho$ (that is, the nucleation rate per unit length of the system) is independent of $L$, and we can represent it as
\begin{equation}
\rho=\frac{\mu_0 \delta}{L_c}\rho_*\,,\;\;\;\mbox{where}\;\;\;\rho_*=R\, e^{-KS_0}\ll 1\, ,
\label{D10}
\end{equation}
and $R$ is a dimensionless pre-factor that depends on dimensionless parameters $K$, $\delta$ and $\mu_0/D_0$.  Importantly, the nucleation problem is mathematically equivalent to the over-damped limit of theory of homogeneous nucleation due to Langer \cite{Langer}, see Appendix C. The theory of Langer corroborates our argument that, for $L\gg L_c$, the nucleation rate is proportional to $L$. Furthermore, his theory makes possible to calculate pre-factor $R$ explicitly. We will not need the pre-factor, however, as we are only interested in the leading-order approximation for $\ln (\mu_0 \delta\,T_e)$.

\begin{figure}[ht]
\hspace{0.1in}\includegraphics[width=2.0 in,clip=]{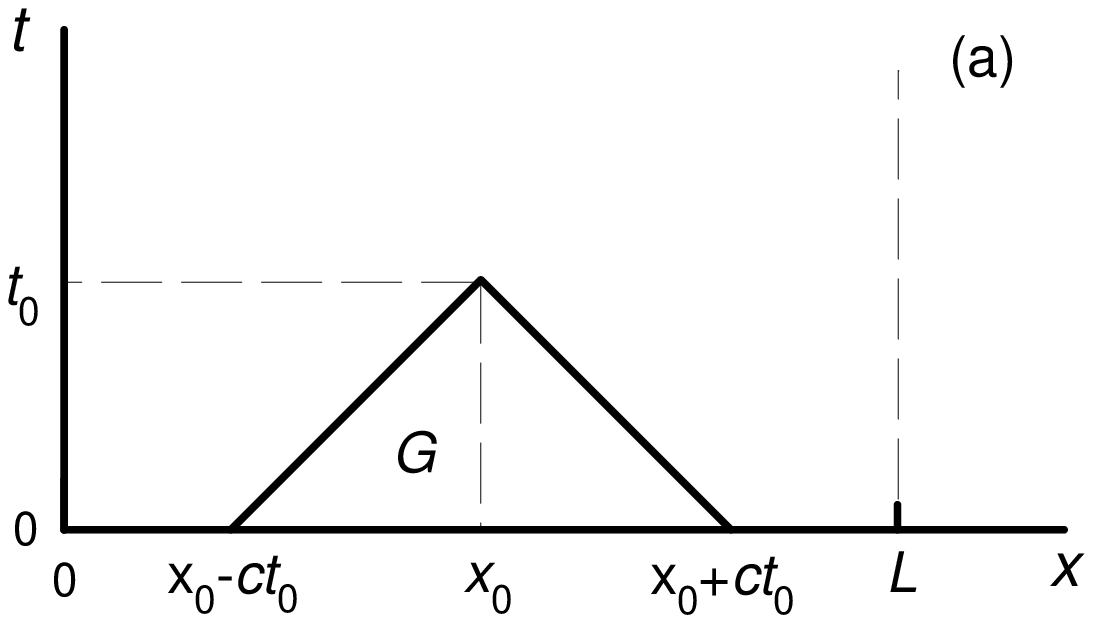}
\includegraphics[width=2.15 in,clip=]{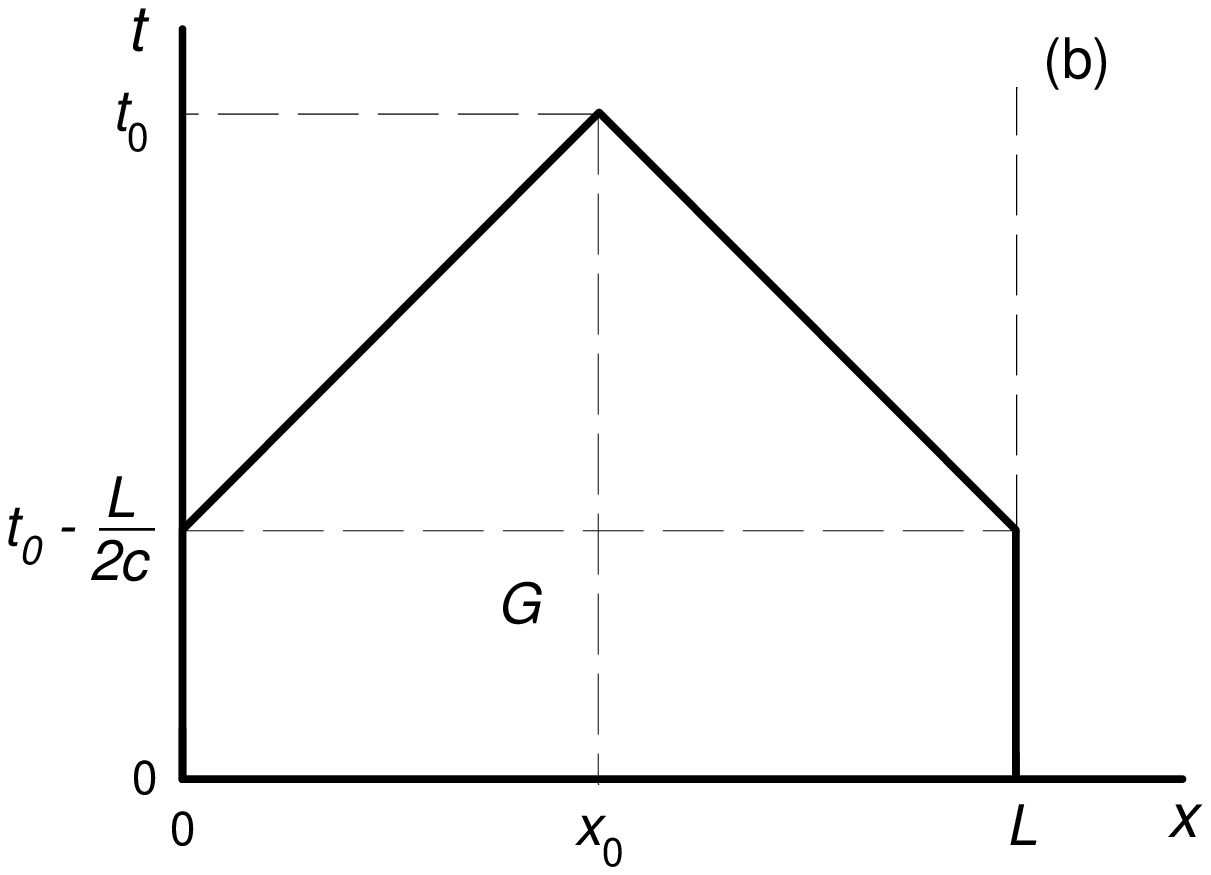}
\caption{Event horizon produced by two extinction fronts: at $2ct_0<L$ (a) and $2ct_0>L$ (b).} \label{horizon}
\end{figure}

With nucleation rate density (\ref{D10}) at hand, we now consider the following problem. Let at $t=0$ the whole system, with $L\gg L_c$, be in the populated state $q_2=1+\delta$. The nucleation rate density $\rho$ is independent of $x$ and $t$. After a critical nucleus (of size $\sim L_c\ll L$) develops, two deterministic extinction fronts form and propagate in both directions with speed  $c\simeq \sqrt{\mu_0 D/2}$, see Eq.~(\ref{speed}). What is the probability ${\cal P}_{q_2}(x_0,t_0)$ to still observe $q=q_2$ at point $x=x_0$ at time $t=t_0>0$? For this to happen, no critical nucleus should have appeared in space-time domain $G$ within the event horizon produced by the two incoming extinction fronts.
Taking into account the finite size $L$ of the system and the periodic boundary conditions, we find that, for $2ct_0<L$, domain $G$ is determined by conditions
\begin{equation}
\left|x- x_0\right|<c(t_0-t)\mbox{~~and~~} 0<t<t_0\, ,
\label{D20}
\end{equation}
whereas for $2ct_0>L$ it is determined by conditions
\begin{eqnarray}
0<x<L&\mbox{~~for~~}& 0<t<t_0-\frac{L}{2c}\, ,\mbox{~~and}
\nonumber\\
\left|x- x_0\right|<c(t_0-t)&\mbox{~~for~~}& t_0-\frac{L}{2c}<t<t_0\,,
\label{D30}
\end{eqnarray}
see Fig.~\ref{horizon}. Indeed, because of the periodic boundary conditions (which bring translational invariance of the problem) we can always choose $x_0=L/2$,
so ${\cal P}_{q_2}(x_0,t_0)$ is actually independent of $x_0$.  The space-time area $\sigma(t_0)$ of domain $G$ is equal to
\begin{equation}
\sigma(t_0)=\left\{
\begin{array}{lc}
ct_0^2&\mbox{~~for~~} 2ct_0 <L\,,\\
Lt_0-\frac{L^2}{4c}&\mbox{~~for~~} 2ct_0 >L\,.
\end{array}
\right.
\label{D40}
\end{equation}
By virtue of the Poisson statistics, we obtain
\begin{equation}
{\cal P}_{q_2}(t)=e^{-\rho \sigma(t)}=\left\{
\begin{array}{lc}
\exp\left(-\rho ct^2\right)&\mbox{~~for~~} 2ct <L\,,\\
\exp\left(-\rho Lt+\frac{\rho L^2}{4c}\right)&\mbox{~~for~~} 2ct >L\,.
\end{array}
\right.
\label{D50}
\end{equation}

\begin{figure}[ht]
\includegraphics[width=2.5 in,clip=]{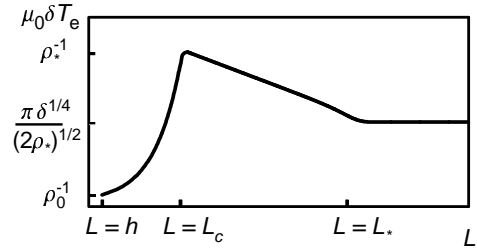}
\caption{Shown, on a log-log scale, is a sketch of system-size dependence~(\ref{D60}) of the rescaled MTE of a population exhibiting a very strong Allee effect.}
\label{long}
\end{figure}

The MTE can be calculated from $T_e=\int_0^{\infty} {\cal P}_{q_2}(t)\,dt$, and we obtain
\begin{equation}\label{D55}
   T_e=\sqrt{\frac{\pi}{4 \rho c}}\,\mbox{erf}\left(\sqrt{\frac{\rho}{c}}\,\frac{L}{2}\right)
   +\frac{1}{\rho L}\,e^{-\frac{\rho L^2}{4c}}\,,
\end{equation}
where $\mbox{erf}(\dots)$ is the error function \cite{Stegun}. Using expression~(\ref{D10})
for $\rho$, we can rewrite Eq.~(\ref{D55}) as
\begin{equation}
\mu_0 \delta T_{e}= \frac{\pi \delta^{1/4}}{(2 \rho_*)^{1/2}}\,\mbox{~erf}\,\left(\frac{L}{L_*}\right)+\frac{1}{\rho_*}\, \frac{L_c}{L}\, e^{-L^2/L_*^2}\,,
\label{D60}
\end{equation}
where the characteristic  length scale $L_*$ (which is exponentially large in $K$) is defined as
\begin{equation}
L_*=\sqrt{\frac{2}{\pi \rho_*}} \,\,\delta^{-1/4}\, L_c=2\delta^{-3/4}\,\sqrt{\frac{\pi D}{\mu_0\rho_*}}\,.
\label{D70}
\end{equation}
For $L_c\ll L\ll L_*$ the first term in Eq.~(\ref{D60}) can be neglected, and we recover, up to a pre-exponent,
the result from subsection \ref{caseB} A: $\mu_0 \delta T_{e} \sim (L_c/L) e^{KS_0}$.  However,
for exponentially large systems, $L\gg L_*$, the second term is negligible, whereas $\mbox{erf}(L/L_*)\to 1$, and we arrive at
$\mu_0 \delta T_e \sim 1/\sqrt{\rho_*} \sim e^{KS_0/2}$. Note that, if we interpret this new asymptote as exponential of some effective WKB action, this action will be
twice as small as the action obtained for $L\ll L_*$.  A sketch of the overall dependence of the MTE on the system length is presented, on a log-log scale, in Fig.~\ref{long}. Evident is a maximum of the MTE at $L$ much larger than $L_c$ but much smaller than $L_*$.

The exponentially large characteristic length scale $L_*$ comes, quite naturally, from the balance between the nucleation rate, $\rho L=\mu_0 \delta \rho_* L/L_c$ and the traverse rate of the deterministic extinction fronts through the system, $\sim c/L$. At $L\ll L_*$ the population is typically going extinct
via formation of only one critical nucleus, whereas at $L\gg L_*$ multiple nucleation acts typically occur.

Finally, the behavior of the MTE versus $L$, depicted in Fig.~\ref{long}, can be understood as follows. At $L_c<L < L_*$ it is the formation of a single
nucleus that serves as a bottleneck of the extinction process. As the formation rate of the nucleus
goes up linearly with $L$,  the logarithm of the MTE goes down linearly with $L$ in this regime. The linear decrease reaches a plateau at $L\gtrsim L_*$ when multiple nucleation acts occur, and multiple extinction fronts are at work.

\section{Discussion}
\label{discussion}
When an isolated stochastic population resides in a refuge of a large but finite size, it ultimately goes extinct with certainty. We have developed WKB approximation to the quasi-stationary multi-variate probability distribution of the population sizes and arrived at an effective Hamiltonian mechanics that encodes  the most probable path the population takes on the way to extinction, and enables one to evaluate the mean time to extinction (MTE). The most general, spatially discrete version of WKB equations employs lattice Hamiltonian (\ref{H}) and is valid for (almost) any relation between the migration rate coefficient $D_0$ and the characteristic rate coefficient $\mu_0$ of the on-site birth-death dynamics. For example, one can use these equations to address an interesting regime, in populations with an Allee effect, where
discreteness of the lattice and a low migration rate conspire to cause propagation failure of deterministic fronts (of either extinction, or colonization), see Ref. \cite{Carpio} and references therein. If the migration is much faster than the on-site population dynamics, it can be described as diffusion, and one arrives at an effective continuous classical mechanics, Eqs.~(\ref{H20})-(\ref{onsiteH}), where one has  to find an activation trajectory: the most probable path of the population to extinction. In the absence of Allee effect (extinction scenario I) and for a very strong Allee effect the most probable path to extinction is an instanton -- a proper heteroclinic connection in the functional phase space of the system.

The extinction dynamics, and the MTE, can be very different depending on whether the population exhibits, or not, Allee effect, as well as on the conditions at the refuge boundaries. The most dramatic differences appear for a sufficiently large system size, $L\gg L_c$. In this case, in the absence of Allee effect (extinction scenario I), the MTE continues to grow exponentially with the system size. When a very strong Allee effect is present, however, extinction occurs via formation of a critical nucleus, and the MTE becomes, up to a pre-exponent, independent of the system size. We have obtained detailed results by assuming that the birth and death rate coefficients are such that the system is close to its characteristic bifurcation (transcritical  or saddle-node in scenarios I and II, respectively). In these cases
one obtains universal Hamiltonians~(\ref{triangle}) and (\ref{parabolic}), describing two broad classes of population models: without Allee effect, and with a very strong Allee effect, respectively. We have also revisited the model system $A\to 2A$ and $2A\to 0$ and shown that, close to the critical system size $L_c$, this system belongs to the universality class described by Hamiltonian~(\ref{triangle}).

For a very strong Allee effect we have mapped  the extinction problem into the over-damped limit of
theory of homogeneous nucleation due to Langer, where the corresponding stationary Fokker-Planck equation is integrable.  This connection gives a natural explanation to the integrability of the zero-energy WKB problem considered in section \ref{caseB}.   In very large systems  the MTE starts to go down with the system size, so there is an optimal refuge size for which the MTE is maximum. At still larger systems
the dependence of the MTE on the system size reaches a plateau. Here multiple nucleation acts occur, and multiple extinction fronts are at work.

For extinction scenario I, the Elgart-Kamenev algorithm of forward and backward iterations \cite{EK1} yields accurate results for the MTE, and for the most probable path to extinction.
An efficient algorithm that would deal with spatial populations exhibiting extinction scenario II is unavailable as of present.  This hinders progress of theory beyond the completely integrable case of a very strong Allee effect. The weak-Allee-effect regime remains \textit{terra incognita}. This includes evaluation of the MTE for periodic or reflecting boundaries, see section \ref{deterministic}C2, where the choice between different possible paths to extinction is not obvious.

In the general part of our derivation, section \ref{master}, we presented the WKB theory for single-step birth-death processes, and in one spatial dimension. A generalization to multiple-step processes (such as a simultaneous birth or death of more than one individual) is straightforward, see section \ref{nonuniversal}, and was already introduced in Ref.~\cite{EK1}.  Higher spatial dimensions can be also taken care of. For scenario I this was observed in Ref.~\cite{EK1}. For scenario II (a very strong Allee effect), the problem remains integrable in higher dimensions, except that the critical nucleus must in general be found numerically.
More challenging generalizations include multiple populations (competition, predation, infection/recovery, \textit{etc.}) and environmental noise.

Our WKB calculations were based on the assumption that the classical action for the \textit{one-site} problem is much greater than unity. This assumption necessitates $K\gg 1$. Our fast-migration results, however, strongly
suggest that this criterion can be relaxed. For example, it is obvious that, for homogeneous-in-space regimes of extinction,  one can treat the whole system as a single site, see Eqs.~(\ref{MTE10}) and (\ref{onesiteB10}), and it is the resulting action for the \textit{whole} system that only needs to be large for WKB theory to hold. As $N\gg 1$, the latter condition can be satisfied even for $K\lesssim 1$. For inhomogeneous-in-space extinction regimes, it should suffice to demand that the action contributed by regions whose spatial dimension is of order of the characteristic diffusion length $l \sim L_c$ be much greater then unity. This necessitates $K L_c/h \sim K (D_0/\mu_0)^{1/2} \gg 1$. For a fast migration, $D_0\gg \mu_0$, this condition is much softer than $K\gg 1$, and it may be be further relaxed close to characteristic bifurcations of the on-site Hamiltonians.

Put in a more general context, this work dealt with rare large fluctuations in spatial stochastic systems far from thermal equilibrium. The last decade has seen a surge of interest in a similar class of problems in the context of steady-state currents in spatial systems of interacting particles, driven by reservoirs at the boundaries, see \textit{e.g.} Ref. \cite{Derrida} and references therein. WKB approximation, bringing about the Hamilton-Jacobi or, alternatively, Hamilton's formalism in a functional phase space, has been instrumental in the analysis of those systems as well \cite{Bertini,Bertini2,Tailleur}.


\subsection*{Acknowledgments}
We gratefully acknowledge extensive discussions with Alex Kamenev. We also thank Bernard Derrida, Vlad Elgart, Omri Gat and Michael Khasin for discussions. This work was supported by the Israel Science
Foundation (Grant No. 408/08), by the
U.S.-Israel Binational Science Foundation (Grant No. 2008075), by the Russian Foundation for Basic Research (Grant No. 10-01-00463),
and by the Lady Davies
Fellowship Trust.

\section*{Appendix A. Absorbing boundaries: governing equations and spatial boundary conditions}
\renewcommand{\theequation}{A\arabic{equation}}
\setcounter{equation}{0}

Here we consider a refuge with absorbing boundaries. The individuals can exit the refuge through its edges $i=1$ and $i=N$ (with the same migration rate coefficient $D_0$ as in the bulk), but no individuals can enter the sites $i=1$ and $i=N$ from outside. In particular, this setting models the extreme situation when the conditions outside of the refuge are so harsh that the individuals die there instantaneously.
In this case master equation (\ref{master1}) needs to be replaced by the following one:
\begin{widetext}
\begin{eqnarray}
\partial_t P(\mathbf{n},t) &=&
\sum_{i=1}^N\lambda(n_i-1) P(\hat{\mathbf{n}},n_i-1,t)+\mu(n_i+1) P(\hat{\mathbf{n}},n_i+1,t) -[\lambda(n_i)+\mu(n_i)] P(\mathbf{n},t)\nonumber \\
 &+&D_0\sum_{i=2}^{N-1} (n_{i-1}+1) P(\hat{\mathbf{n}},n_{i-1}+1,n_i-1,t)+ (n_{i+1}+1) P(\hat{\mathbf{n}},n_i-1,n_{i+1}+1,t) -2 n_i P(\mathbf{n},t)  \nonumber \\
&+&D_0 \left[(n_{1}+1) P(\hat{\mathbf{n}},n_{1}+1,t)+ (n_{2}+1) P(\hat{\mathbf{n}},n_1-1, n_{2}+1,t) -2 n_1 P(\mathbf{n},t)\right.
\nonumber\\
&+&\left.(n_{N-1}+1) P(\hat{\mathbf{n}},n_{N-1}+1,n_N-1,t)+ (n_{N}+1) P(\hat{\mathbf{n}},n_{N}+1,t) -2 n_N P(\mathbf{n},t)\right]\,.
\label{BC10}
\end{eqnarray}
\end{widetext}
Going over to the eigenvalue problem, as in Eq.~(\ref{qsd2}), and applying WKB approximation (\ref{WKB}),  we obtain the following WKB-Hamiltonian:

\begin{widetext}
\begin{eqnarray}
H (\mathbf{q},\mathbf{p}) &=&\mu_0 \sum_{i=1}^{N} \Bigl[\bar{\lambda}(q_i) \left(e^{p_{i}}-1\right) +\bar{\mu}(q_i) \left(e^{-p_{i}}-1\right)\Bigr]
+ D_0 \sum_{i=2}^{N-1} \Bigl[q_{i-1}\left(e^{p_i-p_{i-1}}-1\right)+q_{i+1} \left(e^{p_i-p_{i+1}}-1\right)\Bigr]
\nonumber \\
&+&D_0\left[q_1\left(e^{-p_1}-1\right)+q_2\left(e^{p_1-p_2}-1\right)\right]
+D_0\left[q_{N-1}\left(e^{p_N-p_{N-1}}-1\right)+q_N\left(e^{-p_N}-1\right)\right]\, .
\label{BC20}
\end{eqnarray}
\end{widetext}
This lattice Hamiltonian [\textit{cf.} Eq.~(\ref{H})]
holds
for any relation between $D_0$ and $\mu_0$ \cite{toosmallD}. Now let us consider the limit of $D_0\gg \mu_0$. Here for smooth solutions such as, \textit{e.g.} activation trajectories,  one has $|p_i-p_{i-1}|\ll 1$. Proceeding as in Sec. \ref{master},  we can Taylor-expand the migration part of the Hamiltonian:
\begin{eqnarray}
 \!\!\!\! H_m(\mathbf{q},\mathbf{p}) &=& D_0\sum_{i=2}^N \Big[-\left(q_i-q_{i-1}\right)\left(p_i-p_{i-1}\right) \nonumber \\
  \!\!\!\! &+&\!\!\!\frac{1}{2}(q_{i}+q_{i-1})\left(p_i-p_{i-1}\right)^2 \Big]\nonumber \\
 \!\!\!\!  &+&\!\!\!D_0 \left[q_1 \left(e^{-p_1}-1\right) + q_N \left(e^{-p_N}-1\right)\right].
  \label{H10app}
\end{eqnarray}
Now consider the Hamilton's equation for $dp_1/dt$:
\begin{equation}
\frac{dp_1}{dt}=-D_0\, \left(e^{-p_1}-1+p_2-p_1\right)+ \dots\,,
\label{BC30}
\end{equation}
where $\dots$ denote small corrections coming from the on-site Hamiltonian ${\cal O}(\mu_0)$ and higher-order terms in $p_2-p_1$.  The characteristic time scale of the dynamics of the system (for example, on an activation trajectory) is $\mu_0^{-1}$ (or longer when a bifuraction is approached). Therefore, the left hand side is small, and we obtain, in the leading order in $\mu_0/D_0$,
\begin{equation}\label{e1}
    e^{-p_1}-1+p_2-p_1 \simeq 0 \,.
\end{equation}
As $|p_2-p_1|\ll 1$, the only way to satisfy this condition is to assume that $p_1\ll 1$ which yields, up to small corrections, $p_2=2p_1$. Rewriting this relation as $p_1-(p_2- p_1)=0$ and going over to continuous description,
we obtain
\begin{equation}\label{e2}
    p(x=0,t)-h \,\partial_x p(x=0,t)=0
\end{equation}
or, in the leading order, simply $p(x=0,t)=0$.

Now we consider the Hamilton's equation for $dq_1/dt$. Up to small corrections, we obtain
\begin{equation}
\frac{dq_1}{dt}=D_0\, \left(q_2-2q_1\right)+ \dots\,,
\label{BC40}
\end{equation}
so again $q_2=2q_1+$ small corrections. This yields, in the continuous description,
\begin{equation}\label{e3}
    q(x=0,t)-h \,\partial_x q(x=0,t)=0
\end{equation}
or, in the leading order,
$q(x=0,t)=0$.  Repeating these arguments for site $i=N$ we obtain
\begin{equation}\label{e4}
q(L,t)+h \,\partial_x q(L,t)=0,\;\;\;\;
p(L,t)+h \,\partial_x p(L,t)=0
\end{equation}
\noindent or, in the leading order,  $q(x=L,t)=p(x=L,t)=0$.

When going over to continuous description in the bulk, one
arrives at the same continuous Hamiltonian (\ref{H20})-(\ref{onsiteH}) as in the periodic case.

We note that the gradient terms that appear in Eqs.~(\ref{e2}), (\ref{e3}) and (\ref{e4}) can be legitimately taken into account as small corrections to the zero boundary conditions for $q$ and $p$. Indeed, they are of relative order $(\mu_0/D_0)^{1/2}$ [because the characteristic length scale of the problem is $l\sim (D/\mu_0)^{1/2}=h(D_0/\mu_0)^{1/2}$], whereas the omitted terms -- both in the boundary conditions and in the Hamilton's equations in the bulk -- are much smaller: of order $\mu_0/D_0$.
Close to the transcritical bifurcation, see Sec. \ref{caseA}, one should replace $\mu_0$ by $\mu_0 \delta$ in these estimates.

Zero boundary conditions for the momentum also appear in the context of large deviations in open systems of interacting particles, driven by reservoirs at the boundaries \cite{Bertini,Tailleur}.

\section*{Appendix B}

\renewcommand{\theequation}{B\arabic{equation}}
\setcounter{equation}{0}

The statement that the activation trajectory in extinction scenario I must be a heteroclinic connection AB in the functional phase space $\left\{q(x),p(x)\right\}$  relies on the presence and linear stability properties
of fixed points -- that is, steady-state solutions with specified boundary conditions in space -- of Eqs.~(\ref{p100}) and~(\ref{p110}).   These are described in Appendix B1.
This information is then used in Appendix B2 to prove the statement. Appendix B3 presents a
linear stability analysis of fixed points A, C and D of scenario II.

\subsection*{1. Scenario I: Functional fixed points and their linear stability}

As a typical example of scenario I, we consider universal Hamiltonian (\ref{triangle}) introduced
in section \ref{universal}. There are three zero-energy fixed points here.

\subsubsection*{(i) Fixed point A: $Q=Q_s(x)$, $P=0$}

Here we put $Q(x,t)=Q_s(x)+q(x,t)$ and $P(x,t)=p(x,t)$ and linearize rescaled Eqs.~ (\ref{eqp10}) and (\ref{eqp20}) with respect to  $q$ and $p$. The linearized equations are
\begin{eqnarray}
  \partial_t q  &=& -2Q_s (x)\,q + q + \partial_x^2 q + 2Q_s(x)\,p\,, \label{eqp13}\\
  \partial_t p &=& 2Q_s(x)\,p -p-\partial_x^2 p\,, \label{eqp23}
\end{eqnarray}
subject to zero boundary conditions for $q$ and $p$ at $x=0$ and $x=\tilde{L}>\tilde{L_c}$. Start with Eq.~(\ref{eqp23}) and look for eigenmodes of the form $p(x,t)=e^{E t} \psi(x)$. Eigenfunctions $\psi(x)$ satisfy Schr\"{o}dinger equation
\begin{equation}\label{Sh1}
\psi^{\prime\prime}(x)+[\mathfrak{E}-V(x)] \psi(x)=0\,,
\end{equation}
where $\mathfrak{E}=E+1$, and $V(x)=2 Q_s(x)$.

Now we will prove a simple comparison theorem. Consider an auxiliary equation:
\begin{equation}\label{P1}
\psi^{\prime\prime}(x)+\left[\mathfrak{E}-\frac{1}{2}V(x)\right] \psi(x)=0\, .
\end{equation}
By virtue of Eq.~(\ref{instate}), it has nontrivial solution $\psi(x)\propto Q_s(x)$
at $\mathfrak{E}=1$. This solution obeys zero boundary conditions at $x=0$ and $\tilde{L}$
and has no nodes inside the interval $0<x<\tilde{L}$.
As a result, $\mathfrak{E}=1$ is the lowest eigenvalue of the auxiliary problem.
Now,  our original potential $V(x)$ in Eq.~(\ref{Sh1}) is higher everywhere, except at points $x=0$ and $\tilde{L}$, than auxiliary potential $V(x)/2$. Therefore, the lowest eigenvalue of original problem~(\ref{Sh1}) is strictly greater than the lowest eigenvalue
of the auxiliary problem. Hence $\min \mathfrak{E}>1$, and so \textit{all} eigenvalues $E$ are positive.

Now we turn to Eq.~(\ref{eqp13}). Here $q(x,t)$ is forced by term $2 Q_s(c)\,p(x,t)$. Let us
expand both the forcing and the solution that we are seeking in the complete set of eigenfunctions $\psi_n(x)$ of the momentum:
\begin{equation}\label{exp1}
    2 Q_s(x) p(x,t) = \sum_{n=1}^{\infty} b_n e^{E_n t} \psi_n(x)\,,
\end{equation}
and
\begin{equation}\label{exp2}
q(x,t)=\sum_{n=1}^{\infty} f_n(t) \psi_n(x)\,.
\end{equation}
We obtain equation
$$
\frac{d f_n}{dt}+E_n f_n=b_n e^{E_n t}\,,
$$
for $f_n(t)$, whose general solution is
$$
f_n(t)=a_n e^{-E_n t}+\frac{b_n}{2 E_n}e^{E_n t}\,.
$$
As $E_n>0$ for all $n=1,2,\dots$, we see that deterministic hyper-plane $p=0$ is the stable manifold of fixed point A, as expected from the deterministic theory. The unstable manifold involves a non-zero $p$.

\subsubsection*{(ii) Fixed point B: $Q=0$, $P=P_e(x)$}

Here we put $Q=q(x,t)$ and $P=P_e(x)+p(x,t)$. The linearized equations are
\begin{eqnarray}
  \partial_t q  &=& -2Q_s(x) \,q +q +\partial_x^2 q\,, \label{eqp15}\\
  \partial_t p &=&2Q_s(x) \,p - p -\partial_x^2 p - 2Q_s(x)\, q \,. \label{eqp25}
\end{eqnarray}
where we have used the fact that, for Hamiltonian (\ref{triangle}),  $P_e(x)=-Q_s(x)$. The analysis here
is very similar to that for fixed point A. We first
consider Eq.~(\ref{eqp15}) and look for eigenmodes $q(x,t)=e^{E t} \phi(x)$.
We observe that Eq.~(\ref{eqp15}) for $q$ coincides, up to a sign, with Eq.~(\ref{eqp23}) for $p$. As a result, eigenvalues $E_n$ are ``mirror images" of $E_n$, considered in the context of fixed point A and,
therefore, all of them are negative.

The analysis of forced Eq.~(\ref{eqp25}) for $p$ closely follows that for forced Eq.~(\ref{eqp13}) for $q$. We
expand the forcing and the solution in the complete set of eigenfunctions $\phi_n(x)$ of $q$:
\begin{equation}\label{exp3}
    -2 Q_s(x) q(x,t) = -\sum_{n=1}^{\infty} d_n e^{-E_n t} \phi_n(x)\,,
\end{equation}
and
\begin{equation}\label{exp4}
p(x,t)=\sum_{n=1}^{\infty} g_n(t) \phi_n(x)\,,
\end{equation}
and obtain
$$
\frac{d g_n}{dt}-E_n g_n=-d_n e^{-E_n t}\,.
$$
The general solution is
$$
g_n(t)=c_n e^{E_n t}+\frac{d_n}{2 E_n}e^{-E_n t}\,.
$$
As $E_n>0$ for all $n=1,2,\dots$,  hyper-plane $q=0$ is the unstable manifold of fixed point B,
whereas its stable manifold involves $q(x)>0$.

\subsubsection*{(iii) Fixed point C: $Q=P=0$}

Here the linearized equations are
\begin{eqnarray}
  \partial_t q  &=& q+\partial_x^2 q\,, \label{eqp11}\\
  \partial_t p &=& -p-\partial_x^2 p\,, \label{eqp21}
\end{eqnarray}
and the eigenmodes are elementary:
\begin{eqnarray}
 q &=& A e^{\Gamma_{n_1} t}\, \sin\frac{n_1\pi x}{\tilde{L}}\,,\label{eqp12}\\
  p &=& B e^{\gamma_{n_2} t}\, \sin\frac{n_2\pi x}{\tilde{L}}\,. \label{eqp22}
\end{eqnarray}
where
$\Gamma_{n_1}=1-n_1^2\pi^2/\tilde{L}^2$, $\gamma_{n_2}=-1+n_2^2\pi^2/\tilde{L}^2$, and $n_1,n_2=1,2, \dots$.
We are only interested in solutions with $q(x,t)\geq 0$, therefore the only allowed mode for $q$ is the fundamental: $n_1=1$. This mode, at $\tilde{L}>\tilde{L}_c=\pi$, is unstable. As a result, region $q(x)>0$ in a vicinity of fixed point C belongs to the unstable manifold of this fixed point.

\subsection*{2. Scenario I: Activation trajectories are heteroclinic connections}

Consider extinction scenario I in a system with absorbing boundaries at $x=0$ and $x=L>L_c$.
As one can see from Eqs.~(\ref{p100}) and~(\ref{p110}), hyper-planes $q(x)=0$ and $p(x)=0$ are
invariant manifolds. Each of them is embedded into zero-energy hyper-surface $H\{q(x),p(x)\}=0$.
Therefore, hyper-plane $q(x)=0$ cannot be reached from domain $q(x)>0$ except
via fixed points of Eqs.~(\ref{p100}) and~(\ref{p110}) that belong to hyper-plane $q(x)=0$ (and have finite $p$), or alternatively via  $p(x,t)=-\infty$.  There are exactly two fixed points belonging to hyper-plane $q=0$: B and C. As domain $q > 0$ in a small vicinity of fixed point C belongs to its \textit{unstable} manifold,  fixed point C is
unreachable from domain $q(x)>0$.  On the contrary,  the stable manifold of fixed point B does include $q(x)>0$. Therefore, a trajectory can exist that asymptotically approaches fixed point B at $t\to +\infty$.

Now consider trajectories of Eqs.~(\ref{p100}) and~(\ref{p110}) that come into hyper-plane $q(x)=0$ [at a finite time, and simultaneously at all points of the open interval $(0,L)$] at $p(x,t)=-\infty$. One can show that
$H\{q(x,t),p(x,t)\}>0$ for such trajectories, and so they cannot start from fixed point A. This is similar to what happens in spatially-independent but multi-population systems \cite{KDM}.

Fixed point B has unstable manifold $q(x)=0$, and a stable manifold $\Sigma_B$ that belongs to  domain $q(x)>0$. Each of the two manifolds is $N$-dimensional in the lattice formulation and is embedded into zero-energy hyper-surface $H\{q(x),p(x)\}=0$. Now we see that we need to find a trajectory going from fixed point $A$ to fixed point $B$.  This trajectory must belong
to both hyper-surfaces $\Sigma_A$ and $\Sigma_B$.
In the lattice formulation, each of the two hyper-surfaces is $N$-dimensional
[and is embedded into $(2N-1)$-dimensional hyper-surface $H\{q(x),p(x)\}=0$].
Therefore, hyper-surfaces $\Sigma_A$ and $\Sigma_B$
can intersect, in general, only along a finite set of \textit{one-dimensional} curves
which are trajectories generated by Eqs.~(\ref{p100}) and~(\ref{p110}). These are heteroclinic connections. If there are more than one such connections,  the one with the minimum action along it determines the MTE.

We observed numerically that there is exactly one heteroclinic connection AB in two different examples: for universal Hamiltonian (\ref{triangle}) and for the set of reactions $A\to 2A$ and $2A\to 0$, see section \ref{caseA}. This property apparently holds in a broad class of single-population systems exhibiting extinction scenario I.

\subsection*{3. Scenario II: fixed points A, C and D and their linear stability}

In the limit of a very strong Allee effect  we only need to investigate
the linear stability properties of fixed points A, C and  and D.

\subsubsection*{(i) Fixed points A and C: $q=q_2$ or $0$, $p=0$.}
Linearizing Eqs.~(\ref{p100}) and~(\ref{p110}) around fixed point A, we obtain
\begin{eqnarray}
  \partial_t \delta q  &=& \mu_0 f^{\prime}(q_2) \delta q+D\partial_x^2 \delta q\,, \label{eqp1100}\\
  \partial_t p &=& -\mu_0 f^{\prime}(q_2) p-D\partial_x^2 p\,. \label{eqp2100}
\end{eqnarray}
As $f^{\prime}(q_2)<0$, one can see that deterministic hyper-plane $p=0$ is a stable manifold of fixed point A, whereas $p \neq 0$ is an unstable manifold. The same results hold for fixed point C.


\subsubsection*{(ii) Fixed point D: $q=q_c (x+const)$, $p=0$.}

For periodic boundaries, there is a one-parameter family of fixed points D corresponding to an arbitrary shift with respect to $x$. Because of this degeneracy, there are two eigenmodes that correspond to zero eigenvalue. One of them is neutrally stable:
\begin{eqnarray}\label{BD020}
q(x)-q_c(x)&=& \mbox{const} \,q_c^\prime(x)\,,\\
p&=&0\,;
\label{BD010}
\end{eqnarray}
it corresponds to an infinitesimal shift in $x$ of the critical nucleus $q=q_c(x)$. The other one is algebraically unstable, as it grows \textit{linearly} in time:
\begin{eqnarray}
\label{BD090}
q(x)-q_c(x)&=&\alpha \, q_c^\prime(x)\, t+\tilde{\psi}(x)\,,\\
p&=&C q_c^\prime(x)\,,
\label{BD080}
\end{eqnarray}
where $\tilde{\psi}(x)$ obeys the periodic boundary conditions. We skip here the exact form of function $\tilde{\psi}(x)$, as well as the expression for non-zero constant $\alpha$.

As a result, in the lattice formulation we would have an $(N-1)$-dimensional
stable manifold $\Sigma_D$ that contains point $D$ and leaves hyper-plane $p(x)=0$;
a one-dimensional neutrally stable manifold, belonging to hyper-plane $p(x)=0$,
and an $N$-dimensional unstable manifold with one of its tangent vectors belonging to hyperplane $p(x)=0$. The neutral manifold corresponds to a one-dimensional line of fixed points $D$, parameterized by the exact position of critical nucleus on interval $(0,L)$.

\section*{Appendix C. Fokker-Planck equations for scenarios I and II}
\renewcommand{\theequation}{C\arabic{equation}}
\setcounter{equation}{0}
The Fokker-Planck approximation is a commonly used large-population-size approximation to the master equation \cite{Gardiner}.  Unfortunately, it can only give accurate results for the MTE when the system is sufficiently close to bifurcations describing emergence of established populations (for spatially-independent problems this was observed in Ref.~\cite{Doering}).  Indeed, only in this case the probability distribution of the population size is a slow varying function of the populations size at all relevant population sizes, so that the (truncated) van Kampen system-size expansion \cite{Gardiner} becomes accurate. Here we derive  (lattice versions of) Fokker-Planck equations close to the characteristic bifurcations of extinction scenario I and, for a strong Allee effect, scenario II. These systems are analyzed, in WKB approximation, in sections \ref{universal} and \ref{caseB}, respectively. We also point out to mathematical equivalence between the problem of population extinction for a very strong Allee effect and the over-damped limit of theory of homogeneous nucleation due to Langer \cite{Langer}.

We start from a formal truncated Taylor expansion of the multivariate probability distribution in time-dependent master equation (\ref{master1}). The migration terms in Eq.~(\ref{master1})  become
\begin{widetext}
\begin{eqnarray}
  &&D_0\sum\limits_{i=1}^{N} \Big\{\left(n_{i-1}+1\right)\left[1+\frac{\partial}{\partial n_{i-1}}-\frac{\partial}{\partial n_{i}} +
\frac{1}{2}\left(\frac{\partial}{\partial n_{i-1}}-\frac{\partial}{\partial n_{i}}\right)^2\right]\, P \nonumber \\
&&+ \left(n_{i+1}+1\right)\left[1-\frac{\partial}{\partial n_{i}}+\frac{\partial}{\partial n_{i+1}} + \frac{1}{2}\left(\frac{\partial}{\partial n_{i}}-\frac{\partial}{\partial n_{i+1}}\right)^2\right]\, P
- 2n_i\, P\Big\} \nonumber \\
  &&= D_0\sum\limits_{i=1}^{N} \Big\{2P + \left(n_i+1\right)\left(\frac{\partial}{\partial n_{i}}-\frac{\partial}{\partial n_{i+1}}\right)\, P
+\frac{1}{2} \left(n_i+1\right)\left(\frac{\partial}{\partial n_{i}}-\frac{\partial}{\partial n_{i+1}}\right)^2\, P \nonumber \\
&&+ \left(n_{i+1}+1\right)\left(\frac{\partial}{\partial n_{i+1}}-\frac{\partial}{\partial n_{i}}\right)\, P \Big.
+\,\Big.\frac{1}{2} \left(n_{i+1}+1\right)\left(\frac{\partial}{\partial n_{i}}-\frac{\partial}{\partial n_{i+1}}\right)^2\, P\Big\}\nonumber\\
&&\simeq D_0 \sum\limits_{i=1}^N \left[2P +\left(n_{i}-n_{i+1}\right)\left(\frac{\partial}{\partial n_{i}}-\frac{\partial}{\partial n_{i+1}}\right)\, P
+\frac{n_{i}+n_{i+1}}{2}\left(\frac{\partial}{\partial n_{i}}-\frac{\partial}{\partial n_{i+1}}\right)^2\, P\right]\,.
\label{ps050}
\end{eqnarray}
\end{widetext}
Here we have assumed, for concreteness, periodic boundary conditions. Now we also expand the on-site terms, employ Eq.~(\ref{rates}), go over from $n_i$ to $q_i=n_i/K$ and put everything together. The result is a formal Fokker-Planck equation
\begin{widetext}
\begin{eqnarray}
\partial_t P(\mathbf{q},t)&=&-\mu_0\sum\limits_{i=1}^N\frac{\partial}{\partial q_{i}}\, \left\{\left[\bar{\lambda}(q_i)-\bar{\mu}(q_i)\right]\, P-\frac{1}{2K}\, \frac{\partial}{\partial q_{i}}\,\left[\bar{\lambda}(q_i)+\bar{\mu}(q_i)\right]\, P\right\}\nonumber \\
&+&D_0\sum\limits_{i=1}^N \left(\frac{\partial}{\partial q_{i}}-\frac{\partial}{\partial q_{i+1}}\right)\left[\left(q_{i}-q_{i+1}\right)\, P
+\frac{q_{i}+q_{i+1}}{2K}\left(\frac{\partial}{\partial q_{i}}-\frac{\partial}{\partial q_{i+1}}\right)\, P\right]\,.
\label{ps070}
\end{eqnarray}
\end{widetext}
Even for $K\gg 1$, this equation is only valid, in general, around the attracting fixed point of the deterministic rate equations, where the long-lived population distribution resides. In addition, this equation can hold, for extinction scenario II, in the region around the repelling fixed point. Importantly, it does become accurate for all population sizes $n_i\gg 1$ when the system  is sufficiently close to a bifurcation corresponding to emergence of established populations.  For the SIS model (scenario I) we can assume $\delta\ll 1$ and obtain, after some algebra,
\begin{widetext}
\begin{equation}
 \frac{\partial P(\mathbf{q},t)}{\partial t}  =  \mu_0\sum_{i=1}^N \Big\{ -\frac{\partial}{\partial q_i} \left[q_i (\delta-q_i)P(\mathbf{q},t) \right] +\frac{1}{K} \frac{\partial^2}{\partial q_i^2}\left[q_i P(\mathbf{q},t)\right]\Big\}
 -D_0\sum\limits_{i=1}^N \frac{\partial}{\partial q_{i}}\left[\left(q_{i-1}-2q_i+q_{i+1}\right) P\right]\,.
 \label{FPA}
\end{equation}
\end{widetext}
where we have rewritten the first-derivative migration term in a divergence form and neglected the second-derivative migration term (as it is of next order in $\delta$). Similarly, for three reactions $A\to 0$ and $2A \rightleftarrows 3A$ (scenario II), we obtain, for $\delta\ll 1$ (a very strong Allee effect):
\begin{widetext}
\begin{equation}
   \frac{\partial P(\mathbf{q},t)}{\partial t}  =  \mu_0\sum_{i=1}^N \Big\{ \frac{\partial}{\partial q_i} \left[(q_i-1+\delta)(q_i-1-\delta) P(\mathbf{q},t) \right] +\frac{2}{K} \frac{\partial^2 P(\mathbf{q},t)}{\partial q_i^2} \Big\}
-D_0\sum\limits_{i=1}^N \frac{\partial}{\partial q_{i}}\left[\left(q_{i-1}-2q_i+q_{i+1}\right) P\right]\,.
\label{FPB}
\end{equation}
\end{widetext}
The neglected second-derivative migration term includes an additional $\delta^2$ factor.
In the fast-migration limit, one can replace the lattice formulation by a continuous one,
arriving at functional Fokker-Planck equations close to the bifurcations of scenarios I and II.

Each of Eqs.~(\ref{FPA}) and~~(\ref{FPA}) can be rewritten as a continuity equation,
\begin{equation}
\frac{\partial P}{\partial t} =-\sum\limits_i\frac{\partial  J_i}{\partial q_i}\,.
\label{Lan010}
\end{equation}
For scenario I (at $\delta \ll 1$) the probability flux is
\begin{equation}
J_i=-\frac{\partial {\cal F}(\mathbf{q})}{\partial q_i}\, P-\frac{\mu_0}{K}\, \frac{\partial}{\partial q_i} \left(q_i P\right)\,,
\label{Lan020}
\end{equation}
with free energy
\begin{equation}
{\cal F}(\mathbf{q})=\sum\limits_i \left[\mu_0\, \left(\frac{q_i^3}{3}-\frac{\delta\, q_i^2}{2}\right)+
\frac{D_0}{2}\left(q_i-q_{i-1}\right)^2\right]\, ,
\label{Lan030}
\end{equation}
For scenario II we have, also at $\delta \ll 1$,
\begin{equation}
J_i=-\frac{\partial {\cal F}(\mathbf{q})}{\partial q_i} P-\frac{2\mu_0}{K}\, \frac{\partial}{\partial q_i} P
\label{Lan040}
\end{equation}
and
\begin{eqnarray}
  {\cal F}(\mathbf{q}) &=&\mu_0\, \sum\limits_i  \left[\frac{\left(q_i-1\right)^3}{3}-\delta^2\, \left(q_i-1\right)\right] \nonumber \\
  &+& \frac{D_0}{2}\sum\limits_i \left(q_i-q_{i-1}\right)^2\,.
  \label{Lan050}
\end{eqnarray}
Fokker-Planck Eq.~(\ref{Lan040}) is closely related to the Fokker-Planck equation that appears in the homogeneous nucleation theory of Langer, see Eqs.~(2.15) and~(2.16) of his paper \cite{Langer}. This close relation turns into a full equivalence if one  goes, in Langer's equations,  to the overdamped limit, $A_{ij}=0$, sets $\Gamma=kT=2\mu_0/K$, and specifies free energy ${\cal F}(\mathbf{q})$ as in our Eq.~(\ref{Lan050}). It is crucial that the multi-dimensional effective force, that appears in probability flux ~(\ref{Lan040}), is potential, whereas the diffusion
coefficient $2\mu_0/K$  is $q$-independent, as if coming from \textit{additive} white Gaussian noise in the equivalent Langevin formulation of the problem. In this case the multi-dimensional Fokker-Planck equation is integrable for the purpose of calculating the stationary distribution, and this integrability is closely related to  detailed balance property, see \textit{e.g.} Ref. \cite{Gardiner}.  For a fast migration one can go to the continuous limit and rewrite free energy~(\ref{Lan050}), upon rescaling, as our Eq.~(\ref{freenergy}). The corresponding infinite-dimensional problem remains  integrable for the purpose of finding the stationary solution of the (functional) Fokker-Planck equation. More precisely, the conservation law $F(x,t)=0$, that appears in section \ref{single}, immediately follows, in WKB approximation, from the continuous version of zero probability flux condition $J_i=0$ that solves Eq.~(\ref{Lan010}).  This clarifies the reason behind integrability of the zero-energy WKB problem, considered in section \ref{single} in the context of a very strong Allee effect (and, in Ref. \cite{EK1}, in the context of population explosion).

The situation is less fortunate for extinction scenario I. Although the multi-dimensional force, entering probability flux~(\ref{Lan020}), remains potential, the diffusion coefficient here is $q$-dependent, as if coming from \textit{multiplicative} white Gaussian noise in the Langevin formulation. As a result, the problem of finding the stationary distribution is non-integrable here, and this manifests itself in the non-integrability of the zero-energy WKB problem
considered in section \ref{universal}.

The formal equivalence  between Eqs.~(\ref{Lan010}), (\ref{Lan040}) and (\ref{Lan050}) for a very strong Allee effect  and the equations of theory of homogeneous nucleation  makes it possible to go beyond the leading WKB-order and calculate the sub-leading correction (a pre-exponential factor) to the nucleation rate \cite{Langer}.


\begin{thebibliography}{99}
\bibitem{periodic} For periodic boundary conditions, an $x$-dependent solution can only be unique up to an arbitrary shift in $x$.
\end{thebibliography}
\end{document}